\documentclass[twocolumn,showpacs,amsmath,amssymb]{revtex4}
\usepackage{graphicx}
\usepackage{dcolumn}
\usepackage{bm}
\def\psizel{8.0cm}
\def\psizes{4.0cm}

\begin{document}
\title{Numerical study of drying process and columnar fracture process in granules-water mixtures}
\author{Akihiro Nishimoto}
\email{nisimoto@ton.scphys.kyoto-u.ac.jp}
\affiliation{Department of Physics, Kyoto University, Kyoto 606-8502, Japan}
\author{Tsuyoshi Mizuguchi}
\affiliation{Department of Mathematical Sciences, Osaka Prefecture University, Sakai 599-8531, Japan}
\author{So Kitsunezaki}
\affiliation{Graduate School of Human Culture, Nara Women's University, Nara 630-8506, Japan}

\date{\today}
\begin{abstract}
The formation of three-dimensional prismatic cracks in the drying process of starch-water mixtures is 
investigated numerically.
We assume that the mixture is an elastic porous medium
which possesses a stress field and a water content field.
The evolution of both fields are represented by a spring network
and a phenomenological model with the water potential, respectively.
We find that the water content distribution has a propagating front
which is not explained by a simple diffusion process.
The prismatic structure of cracks driven by the water content field is observed.
The depth dependence and the coarsening process of the columnar structure are also studied.
The particle diameter dependence of the scale of the columns
and the effect of the crack networks on the dynamics of the water content field are
also discussed.
\end{abstract}

\pacs{62.20.Mk,\ 46.50.+a,\ 45.70.Qj,\ 47.56.+r}
\maketitle

\section{Introduction}
Crack patterns are observed in everyday life \cite{Walker}.
Columnar joint in cooled lava, especially, is an intriguing phenomenon
and has fascinated many people for centuries
and has been studied by field works \cite{Folley,Mallet,Peck68,Aydin}.
Theoretical studies have been carried out
from the viewpoint of 
the analysis of the thermal field \cite{Reiter,Degraff93,Grossenbacher},
the ordering of patterns \cite{Budkewitsch,Jagla1},
and size selection of columns by the finite element method \cite{Saliba}.
However, many questions remain open,
such as, `why are the prismatic structures with the polygonal cross sections formed?',
`how are the size of columns and the distribution of polygon types determined?'.
The scale of columns is $0.1 \sim 1\ {\rm m}$ in diameter,
hence it is too hard to study the fracture process
of cooled lava experimentally.

Recently, the patterns of cracks similar to columnar joint are investigated
in the drying process of starch-water mixtures
\cite{Muller1,Muller2,guchi01,Toramaru,Goehring,guchi05,Goehring06}.
Well-controlled experiments are possible 
due to the small scale of columns ($1 \sim 10\ {\rm mm}$ in diameter).
In typical experiments of these studies,
after the mixtures of starch and water are poured into a container,
water evaporates from the sample surface.
During the drying process there are three stages \cite{Muller2,Toramaru,guchi05,Goehring06}.
{\it Stage 1}: The water content decreases uniformly.
{\it Stage 2}: Cracks are formed to extend to the bottom of the sample
and a sudden change of water content occurs.
These cracks have a uniform structure along the vertical direction
and are referred as the primary (type I) cracks \cite{Groisman,Kitsune,Muller2,guchi05,type12}.
In the case that the thickness of a sample is sufficiently large,
the primary cracks appear only between the mixture and the side wall of the container.
{\it Stage 3}: The water content decreases slowly 
and the water content distribution becomes non-uniform, i.e., 
it has a ``front'' \cite{guchi05,Goehring06}.
The front propagates inward and the mixture shrinks non-uniformly
and another type of cracks, i.e., the secondary (type II) cracks are formed.
The secondary cracks show three-dimensional prismatic structure.
The size of columns depends on the evaporation rate \cite{Toramaru,Goehring}
and it coarsens with the distance from the sample surface \cite{Goehring,guchi05}.
In this paper, we focus on the secondary cracks.

By a numerical model, Hayakawa \cite{Hayakawa94b} exhibited 
that a regular prismatic crack structure is formed 
by the steady sweep of external field (temperature) with fixed shape. 
Experimental results of the drying starch, however, suggest that 
the shape of the external field (water content) changed temporally.
In order to study the crack pattern of starch-water mixtures,
the drying process and
the fracture process
should be taken into account three-dimensionally.
Combined numerical analysis of both processes, however, has not been performed yet.
The drying process involving all stages, especially, has not been analyzed,
that is mainly because the water transportation in the porous media are complicated process.

The non-uniform shrinkage 
due to the change of the external field
causes directional propagation of cracks,
and the cracks become the boundary condition of the external field, simultaneously.
The changes of the external field can be induced by cracks as the boundary condition,
as suggested by some studies \cite{Yakobson,Boeck,Sorenssen}.
This type of crack is referred as self-driven crack.
It has neither been discussed in detail nor demonstrated numerically
whether this effect is important or not in the case of starch-water mixtures. 
In some studies on
columnar joints \cite{Hardee,Reiter,Degraff89,Degraff93,Budkewitsch}
and desiccation cracks \cite{Allain,guchi05},
the effect of cracks as the boundary condition is considered to play an important role.
On the other hand,
in studies such as
two-dimensional directional fracture of glass strip \cite{YuseSano93,Ronsin97,Ronsin98},
columnar joints \cite{Grossenbacher}
and desiccation cracks \cite{Komatsu97,Muller1,Muller2,Dufresne03,Dufresne06,Goehring06},
this effect is hypothesized not to be important.

In this paper,
our goal is 
a comprehensive description of the water content distribution
and the three-dimensional realization of the crack formation process
driven by the non-uniform shrinkage
during the drying process of starch-water mixtures.

This paper is organized as follows:
in Sec.~II, details of the model are described.
We assume that the mixture is an elastic porous medium 
which has a stress field and a water content field.
The evolution of the former is represented by a spring network model and
the latter is represented by a phenomenological model with the water potential.
Here we suppose that cracks do not affect the evolution of
the water content distribution.
In Sec.~III, 
numerical results are presented.
A propagating  front of water content distribution
and the prismatic pattern of cracks driven by the water content field are obtained, 
which well capture those observed in the experiments.
The particle diameter dependence of the scale of columns is suggested.
In Sec.~IV, 
the properties of the drying front 
and the validity of some assumptions used in our model are discussed.
The effect of the crack networks on the dynamics of the water content field
and the interaction between the stress field and the water content field
are considered.
The geological columnar joints and 
the characterization of the crack patterns are discussed.
Finally, we conclude this paper with a summary in Sec.~V.

%
%
\section{Model}
\begin{figure}\begin{center}
\includegraphics[width=\psizel]{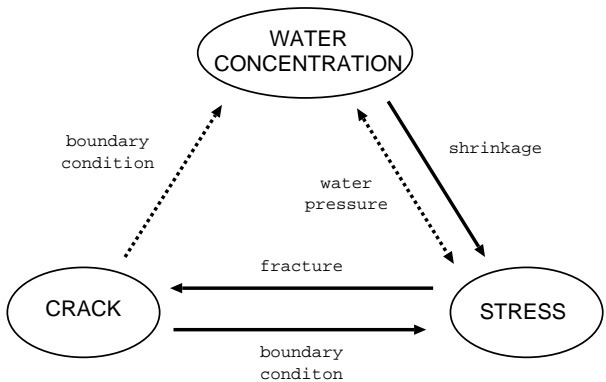}
\caption{
Schematic diagram of the model.
Three elements and interactions between them are drawn.
The interactions indicated by solid arrows are taken into account in our model
while the dashed ones are ignored.
}
\label{model}
\end{center}\end{figure}

The drying fracture process consists of various factors.
Here, as key factors, we focus on the following quantities and processes (Fig.~\ref{model}) :
the change of water content 
by the evaporation and transportation of liquid water and vapor in porous media,
the non-uniform volumetric contraction and the stress intensification,
the formation of cracks,
and the change of the boundary condition of the stress field.

Starch-water mixtures are elastic porous media
composed of solid phase (granules), liquid phase (water) and gas phase (air and vapor).
As the water content decreases by drying,
the state of the mixture changes
from ``capillary fringe'' (liquid is saturated under negative pressure)
to ``funicular'' (liquid is unsaturated and contiguous),
and finally to ``pendular'' (liquid is isolated) \cite{Lu}.
This drying process is mainly driven by the transportation of liquid water and vapor,
and the phase change between them.
In this paper, we adopt the water potential \cite{suction} as one of the key field variables.

For the simplicity, we assume that
the interactions between the crack network and the water content field,
the influences of the water pressure on the stress field,
and the influences of the stress field on the water potential,
indicated by dashed lines in Fig.~\ref{model},
are negligible.
Our numerical simulations, however, exhibit good realizations of the experimental results
as shown in the next section.
The validity of these assumptions are discussed in Sec.~IV.

The stress field and its boundary are represented by the network of springs.
And the water content field and its evolution are represented 
by the equations for the water potential.
The details of the model are stated in the following subsections, i.e.,
the stress field, the water content field using the water potential,
and the interaction between them and their evolution.

\subsection{Stress field}
We construct a cubic lattice with nearest-neighbor (n.n.) and next-nearest-neighbor (n.n.n.) interactions
in order to calculate the stress field \cite{Hayakawa94b}. 
Each interaction is represented by a Hookean spring.
The lattice constant is $a_0$ and
the size of the system is $L_x\times L_y\times L_z$.
The lattice point is expressed as $\alpha\equiv(X,Y,Z)$,
where 
$0\le X\le N_x$,\ $0\le Y\le N_y$,\ $0\le Z\le N_z$,\ 
$L_x=N_x a_0$,\ $L_y=N_y a_0$,\ $L_z=N_z a_0$.
The $z$ axis is chosen as the depth,
the top\ (bottom) is set to $z=0$\ ($z=L_z$).

Let the spring constants for the nearest-neighbor and next-nearest-neighbor interactions
be $k_1$ and $k_2$, respectively.
Then we obtain the spring network of the elastic tensor $C_{ijkl}$
with the components \cite{Hayakawa94b}:
\begin{eqnarray}
C_{iiii}=\frac{k_1+2k_2}{a_0},\\
C_{ijij}=C_{ijji}=C_{iijj}=\frac{k_2}{a_0}.
\end{eqnarray}

We assume that the water content field changes very slowly and
the stress field relaxes much faster than the water content field.
So the stress field is assumed to be balanced at each moment
under conditions induced by the water content field and the boundary condition at that time.
We assume that each spring is broken at a certain critical force, i.e.,
if the force given on a spring exceeds a constant $F_c$,
the spring constant is set to be zero irreversibly.
Using a position vector ${\bf r}_{\alpha}$ of the lattice point $\alpha$,
the force acting on the lattice point $\alpha$ is represented by
\begin{eqnarray}
&&{\bf F}_\alpha=\sum_{\beta\in {\rm n.n.}} K^{\alpha\beta}_1 
\frac{{\bf r}_\alpha - {\bf r}_\beta}{\left|{\bf r}_\alpha - {\bf r}_\beta \right|} 
\left( a_{\alpha \beta}-\left|{\bf r}_\alpha - {\bf r}_\beta \right| \right) \nonumber\\
&&+\sum_{\beta\in {\rm n.n.n.}} K^{\alpha\beta}_2 
\frac{{\bf r}_\alpha - {\bf r}_\beta}{\left|{\bf r}_\alpha - {\bf r}_\beta \right|} 
\left( a_{\alpha \beta}-\left|{\bf r}_\alpha - {\bf r}_\beta \right| \right),\\
&&K^{\alpha\beta}_i=\left\{
\begin{array}{ll}
k_i & ({\rm connected},\ i=1,2)\\
0 & ({\rm broken})
\end{array}
\right.,
\end{eqnarray}
where $a_{\alpha \beta}$ is the natural length of the spring
which connects lattice points $\alpha$ and $\beta$.
The equilibrium condition of the spring network, ${\bf F}_\alpha=0$, is equivalent
to the condition that the elastic energy 
\begin{equation}
E=\sum_{\alpha} E_{\alpha},
\end{equation}
where
\begin{eqnarray}
E_{\alpha}=&&\sum_{\beta\in {\rm n.n.}} 
\frac{1}{2} K^{\alpha\beta}_1 
\left( a_{\alpha \beta}-\left|{\bf r}_\alpha - {\bf r}_\beta \right| \right)^2 \nonumber\\
+&&\sum_{\beta\in {\rm n.n.n.}} 
\frac{1}{2} K^{\alpha\beta}_2 
\left( a_{\alpha \beta}-\left|{\bf r}_\alpha - {\bf r}_\beta \right|\right)^2, \label{eqnEne}
\end{eqnarray}
is minimal under the given boundary condition and the given natural length of each spring.
In our simulations, the free boundary condition is imposed.
Because no friction acts on the boundary of the system, the primary cracks do not appear.

\subsection{Water content field}
The evolution of the water content in unsaturated porous media
consists of various processes.
Here, we use the Campbell's desiccation model \cite{Campbell}
that includes the transportation of liquid water and vapor and the evaporation.
By assuming that the temperature is constant and the speed of evaporation is slow,
the heat of evaporation is ignored.
The evolution equation is described by one variable
because both the liquid water content and the density of vapor are determined from
the water potential as described below.

The effect of the deformation of the spring network 
on the water content field is ignored
and the evolution equation is solved
on the undeformed lattice in the Cartesian coordinates system
because the change of the total volume is smaller than that of water volume \cite{Muller2}.
The change of the boundaries due to the formation of cracks is also ignored.
Note that there are many field variables and constants in this model. 
In order to distinguish them all the field variables are expressed 
by the symbols with an argument of position ${\bf r}$ (or $z$) and time $t$. 

The water potential is defined as the chemical potential per unit mass of water in the mixtures,
compared to that of pure, free water \cite{Campbell,Lu,Iwata}.
The potential is generally negative for unsaturated mixtures.
Although this potential is the sum of several components,
some components are negligible.
The effect of the gravity are negligible
because the size of a starch-water specimen in the experiments is on the order of $1\ {\rm cm}$.
We carried out the experiment \cite{guchi-u}
to investigate the effect of gravity
and confirmed that the columns normal to the free surface are formed
even when the desiccated starch-water mixture in a container is laid on its side.
We assume that there is no external pressure and the water in mixtures is pure,
so we can ignore the overburden potential and the osmotic potential.
Hence, the water potential is well represented by the {\it matric potential} $\psi_m({\bf r},t)$.
%
The matric potential is defined as the amount of work per unit mass of water,
required to transport an infinitesimal quantity of water
from the mixture to a reference pool of water
at the same elevation, pressure and temperature.
The capillary water between particles and the water on the particle surface
contribute to the matric potential.
%
Below we use the matric potential as the water potential, i.e., 
the chemical potential per unit mass of water in mixtures.
The contribution of the capillary water to the potential is expressed as
\begin{equation}
\frac{P_{cap}}{\rho_L},
\end{equation}
where $P_{cap}$ is the capillary pressure, expressed as $-2\sigma/r_{cap}$, 
$\sigma$ is the surface tension of water and $r_{cap}$ is 
the inverse of the mean curvature of the capillary water in the mixture.

From the coexistence condition of the liquid phase and the gas phase,
the relative humidity $h({\bf r},t)$ is related to the matric potential $\psi_m({\bf r},t)$
by the Kelvin eq.,
\begin{equation}
h({\bf r},t) = \exp\left( \frac{M_w}{RT}\psi_m({\bf r},t) \right), \label{Kelvin}
\end{equation}
where $M_w$ is the molecular weight of water,
$R$ is the gas constant
and $T$ is the absolute temperature.

Let $\theta_L({\bf r},t)$ denote the volumetric water content.
Assuming that $\theta_L({\bf r},t)=\theta_S$ in the saturated state
($\theta_S$ corresponds to the porosity of the mixture),
the volumetric gas content $\theta_G({\bf r},t)$ is given as
\begin{equation}
\theta_G({\bf r},t)=\theta_S-\theta_L({\bf r},t). \label{eqnG}
\end{equation}
During the drying process, $\theta_L({\bf r},t)$ decreases as $\theta_G({\bf r},t)$ increases.
$\theta_S$ is assumed to be constant
because we ignore the feedback 
from the stress field (fracture and deformation) to the water content field
and the change of the pore volume due to the change of the water content.

Expressing the fluxes of liquid water and vapor as $J_L({\bf r},t)$ and $J_V({\bf r},t)$
and the evaporation rate per unit volume as $W_{eva}({\bf r},t)$,
the transportation equation of liquid water and vapor is given by
\begin{eqnarray}
\frac{\partial}{\partial t} (\rho_L \theta_L({\bf r},t))
= - \nabla\cdot J_L({\bf r},t) - W_{eva}({\bf r},t), \label{eqnW}\\ 
\frac{\partial}{\partial t} (\rho_V({\bf r},t) \theta_G({\bf r},t))
= - \nabla\cdot J_V({\bf r},t) + W_{eva}({\bf r},t), \label{eqnV}
\end{eqnarray}
where $\rho_L$ is the density of liquid water,
$\rho_V({\bf r},t)$ is the density of vapor.
The incompressibility of water is assumed, i.e., $\rho_L$ is constant.

\begin{figure}
\includegraphics[width=\psizel]{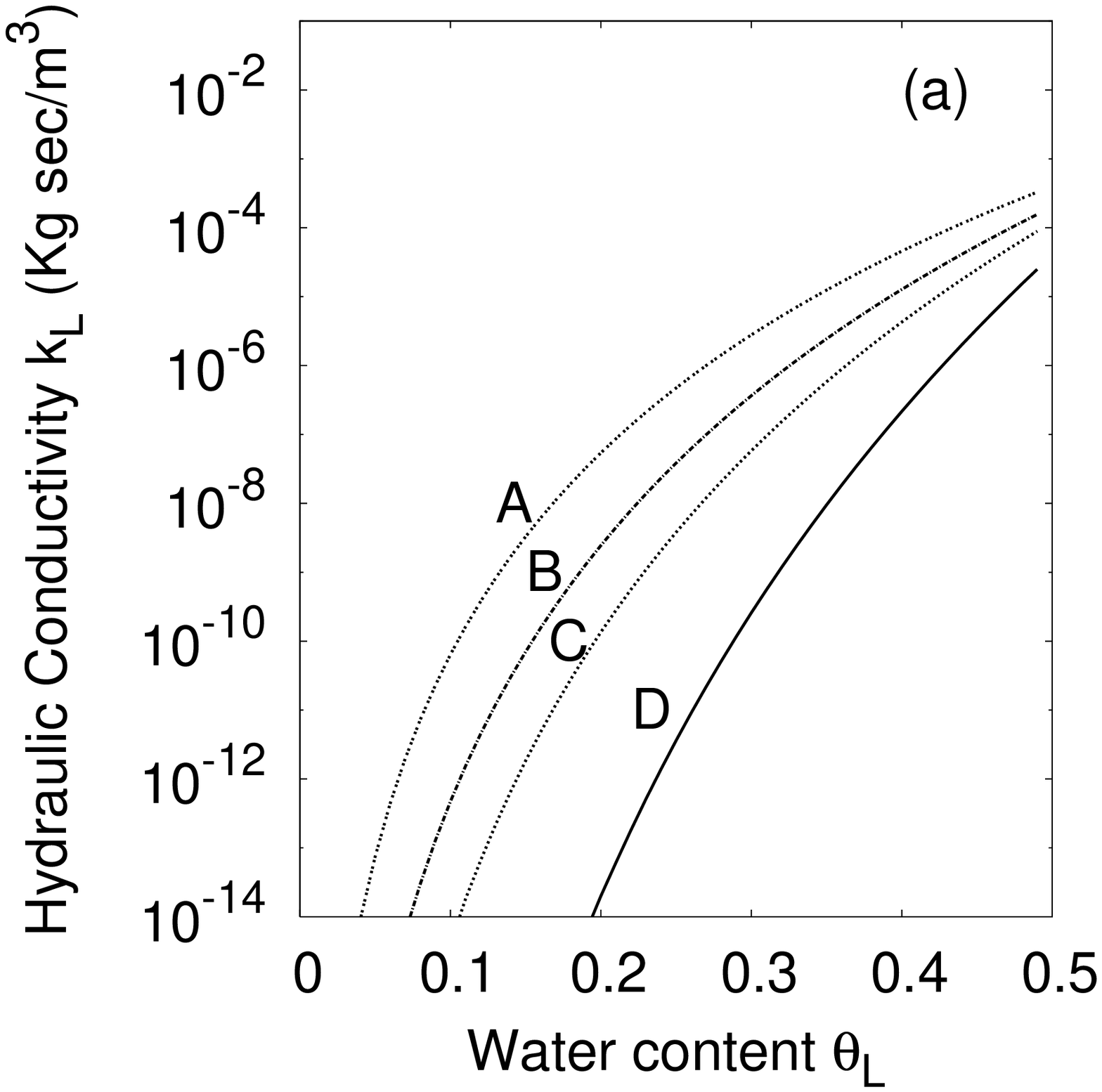}\\
\includegraphics[width=\psizel]{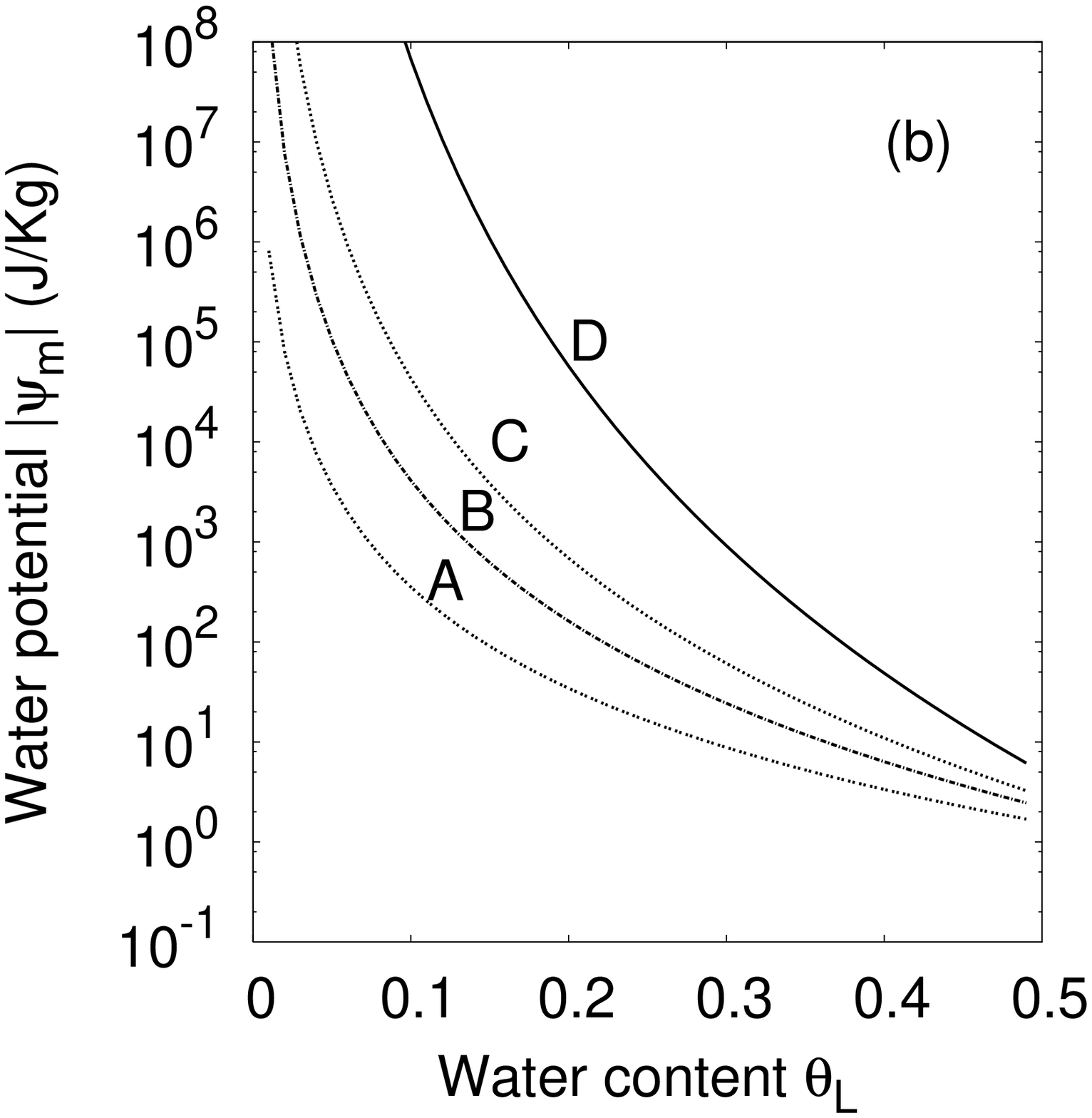}
\caption{Dependence of (a) $k_L$ and (b) $|\psi_m|$ on $d_g$.
A:\ $d_g=0.100\ {\rm mm}$,\ B:\ $d_g=0.050\ {\rm mm}$,\ 
C:\ $d_g=0.030\ {\rm mm}$\ and\ D:\ $d_g=0.010\ {\rm mm}$.}
\label{water_param}
\end{figure}

In the unsaturated porous media, the flux of water is assumed to be 
proportional to the gradient of the water potential (Darcy's law) \cite{Campbell,Lu,Iwata},
\begin{equation}
J_L({\bf r},t) = -k_L({\bf r},t)\nabla \psi_m({\bf r},t),
\end{equation}
where $k_L({\bf r},t)$ is the hydraulic conductivity.
The properties of starch-water mixtures in drying process,
such as the dependences of 
the hydraulic conductivity $k_L({\bf r},t)$ and the water potential $\psi_m({\bf r},t)$ 
on the volumetric water content $\theta_L({\bf r},t)$
have not investigated in detail \cite{Komatsu03,Komatsu01}.
Here, we suppose that the properties of starch-water mixtures are
similar to those of soil-water systems and employ the relations 
confirmed by experiments in the soil mechanics \cite{Campbell,Lu},
\begin{eqnarray}
k_L({\bf r},t) &=& k_{L0} \left( \frac{\theta_L({\bf r},t)}{\theta_S}\right)^{2b+3}, \label{eqnKl}\\
\psi_m({\bf r},t) &=& \psi_{m0} \left(\frac{\theta_L({\bf r},t)}{\theta_S}\right)^{-b}, \label{eqnPm}
\end{eqnarray}
where
\begin{eqnarray}
\psi_{m0} &=& -0.5 d_g^{-\frac{1}{2}}, \label{eqndg}\\
k_{L0} &=& 1.0 \times 10^{-3} \psi_{m0}^{-2},\\
b &=& -2 \psi_{m0} + 0.2\sigma_g. \label{eqnB}
\end{eqnarray}
$\psi_{m0}$ and $k_{L0}$ are material constants.
The units of $\psi_{m0}$ and $k_{L0}$ are ${\rm J/kg}$ and ${\rm kg\ sec/m^3}$, respectively.
$d_g$ and $\sigma_g$ are 
the geometric mean diameter and standard deviation of the particles, respectively (the unit is mm).
The index $b$ is a phenomenological parameter for soil-water systems.
The relations (\ref{eqnKl}) and (\ref{eqnPm}) are explained 
by some naive theories assuming the power-law
distribution of pore size \cite{Campbell}.
The dependences of $k_L({\bf r},t)$ and $|\psi_m({\bf r},t)|$ 
on the diameter $d_g$ are indicated in Fig.~\ref{water_param}.
Further experimental examinations may be required
to confirm the validity for starch-water mixtures.
We, however, expect that the important feature of this model
is the functional shape of the two relations
$k_L(\theta_L)$ and $\psi_m(\theta_L)$
and that the following results are robust against tiny modifications.

The flux of vapor is described by Fick's law,
\begin{equation}
J_V({\bf r},t) = -D_V \epsilon_V \theta_G({\bf r},t) \nabla\rho_V({\bf r},t),
\end{equation}
where $D_V$ is the diffusion constant of vapor and $\epsilon_V$ is the tortuosity factor.
The density of vapor is given by 
\begin{equation}
\rho_V({\bf r},t)=\rho_{V0}h({\bf r},t), \label{eqnrhoV}
\end{equation}
where $\rho_{V0}$ is the saturation vapor density 
and the relative humidity $h({\bf r},t)$ is related to the matric potential by Eq.~(\ref{Kelvin}).

The evolution equation of $\psi_m({\bf r},t)$ is obtained 
from Eqs.~(\ref{Kelvin})-(\ref{eqnrhoV}):
\begin{equation}
f(\psi_m)\frac{\partial \psi_m({\bf r},t)}{\partial t} 
= \nabla\cdot( k_m({\bf r},t) \nabla\psi_m({\bf r},t) ), \label{eqnP}\\
\end{equation}
where
\begin{eqnarray}
f(\psi_m)&=&
\frac{ (\rho_L-\rho_V({\bf r},t))\theta_S }{ -b\psi_m({\bf r},t) }
\left(\frac{\psi_{m0}}{\psi_m({\bf r},t)}\right)^{\frac{1}{b}} \nonumber\\ 
&&+\frac{M_w \rho_V({\bf r},t) \theta_G({\bf r},t) }{ RT },\\
k_m({\bf r},t)&=&k_L({\bf r},t)+k_V({\bf r},t),\\
k_V({\bf r},t)&=&\frac{D_V \epsilon_V \theta_G({\bf r},t) \rho_V({\bf r},t) M_w}{RT}.
\end{eqnarray}

The boundary condition is that the evaporation occurs only at the top surface,
\begin{eqnarray}
-(k_V+k_L) \left.\frac{\partial \psi_m}{\partial z}\right|_{top} &=& -E_p \frac{h(z=0,t)-h_a}{1-h_a},\\
\left.\frac{\partial \psi_m}{\partial z}\right|_{bottom} &=& 0,\\
\left. {\bf n}\cdot\nabla \psi_m\right|_{side} &=& 0,
\end{eqnarray}
where $E_p$ is the evaporation rate for the state that vapor is saturated on the top surface
and $h_a$ is the atmospheric humidity
and ${\bf n}$ is the normal vector at the side boundary.
Note that the relative humidity at the top surface $h(z=0,t)$ changes, 
so the evaporation rate is not constant \cite{evaporation}.

\subsection{Combination of two fields}
The volume shrinkage occurs due to drying,
which is the effect from the water content field to the stress field.
This effect is incorporated into the model
such that the natural length of each spring is 
an increasing  function of the water content $\theta_L({\bf r},t)$, i.e.,
the natural length $a_{\alpha \beta}$ of the spring between the lattice points $\alpha$ and $\beta$
is determined as
\begin{eqnarray}
a_{\alpha \beta}&=&
a_{0\alpha\beta}
\left\{1+\kappa\left( \frac{{\theta_{L\alpha}}+{\theta_{L\beta}}}{2}-\theta_S\right)\right\}, \label{eqna0}\\
\theta_{L\alpha}&=&\theta_L\left(\frac{L_x}{N_x}X,\frac{L_y}{N_y}Y,\frac{L_z}{N_z}Z,t\right), \label{eqna1}\\
a_{0\alpha\beta}&=&\left\{
\begin{array}{ll}
a_0 & ({\rm n.n.})\\
\sqrt{2}a_0 & ({\rm n.n.n.})
\end{array}
\right.. \label{eqna2}
\end{eqnarray}
The parameters of the stress field,
$k_1$, $k_2$,
the expansion coefficient $\kappa$,
and the critical force for the breaking $F_c$,
are supposed to be constant for the simplicity.
The changes in the stress field
may affect the water content field (or the water potential),
however this effect is assumed to be ignored.

We also assume that evaporation does not occur from the crack surfaces as well as the side boundaries.
Therefore, in our model, 
all the field variables, 
such as the water content field $\theta_L({\bf r},t)$ and $\psi_m({\bf r},t)$
are the function of the depth $z$ and time $t$.
The argument of the field variables are changed 
from $({\bf r},t)$ to $(z,t)$ hereafter.
The drainage effect, i.e., the role of the evaporation through the crack
will be discussed in Sec.~IV.

\subsection{Time evolution}
The initial states in our simulations are prepared so that 
the pore is saturated,
\begin{equation}
\theta_L(z,t=0)=\theta_S,
\end{equation}
and randomly chosen $1\%$ of the springs at the surface($z=0$) are broken.

We repeated the following procedures at each time step:
\begin{enumerate}
\item Time steps by $dt$
and the water potential $\psi_m(z,t)$ is calculated from Eq.~(\ref{eqnP})
and the water content is obtained from Eq.~(\ref{eqnPm}).
\item The natural length of each spring is determined 
from the water content field $\theta_L(z,t)$ by Eqs.~(\ref{eqna0})-(\ref{eqna2}).
\item The equilibrium configuration of the stress field is calculated
so that the elastic energy (\ref{eqnEne})
has the minimal value
by the conjugate gradient method \cite{Recipe} with a tolerance $10^{-4}$.
\item If the force given on a spring exceeds the critical value $F_c$, it breaks, 
i.e., the corresponding spring constant changes to zero.
If more than one spring have loads larger than $F_c$, break one which has the maximum force.
\item The procedures 3 and 4 are repeated until no spring is broken,
and then return to the first procedure.
\end{enumerate}

%
%
\section{Numerical Results}
\subsection{Typical patterns}
\begin{figure}
\includegraphics[width=\psizel]{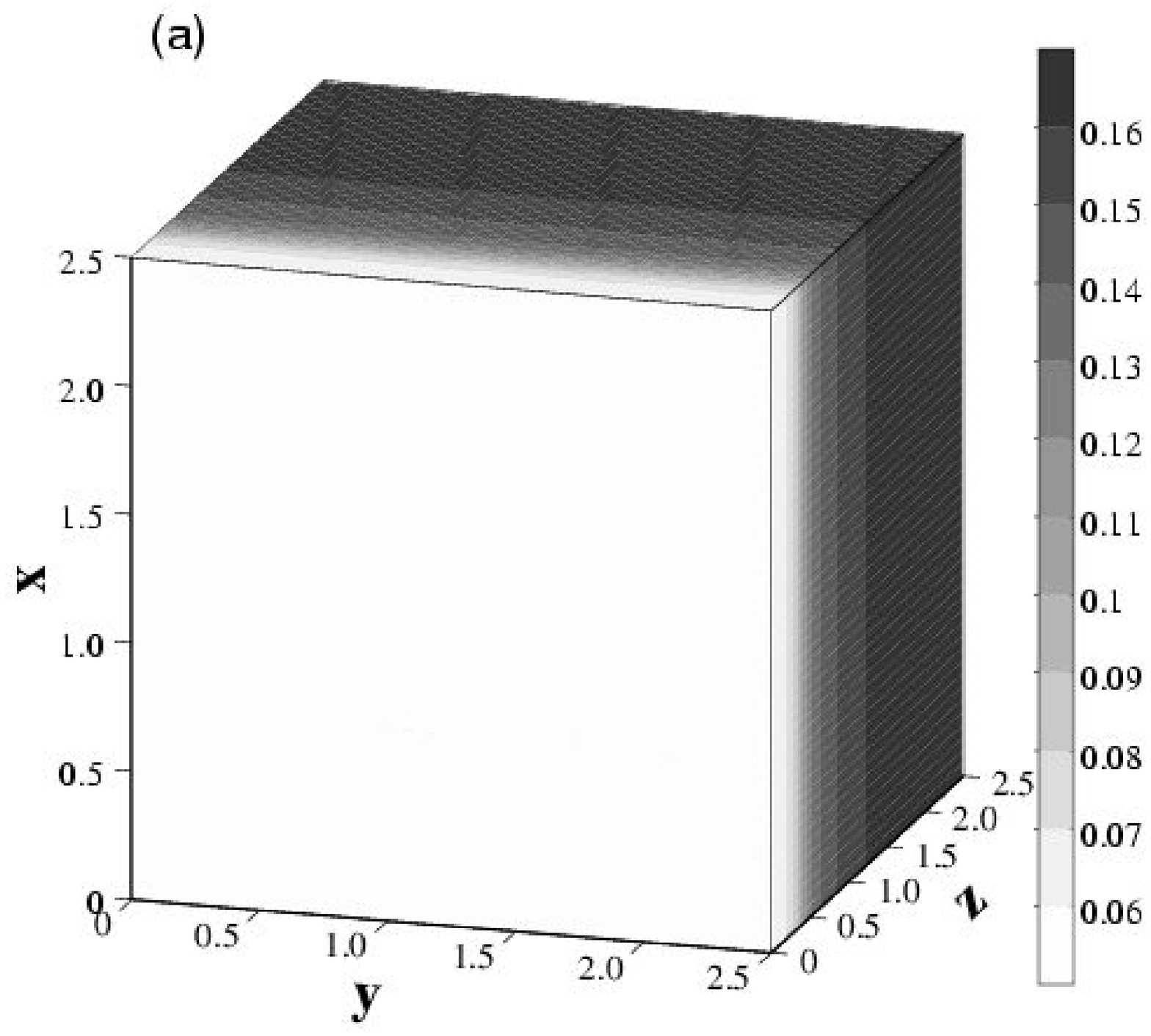}
\includegraphics[width=\psizel]{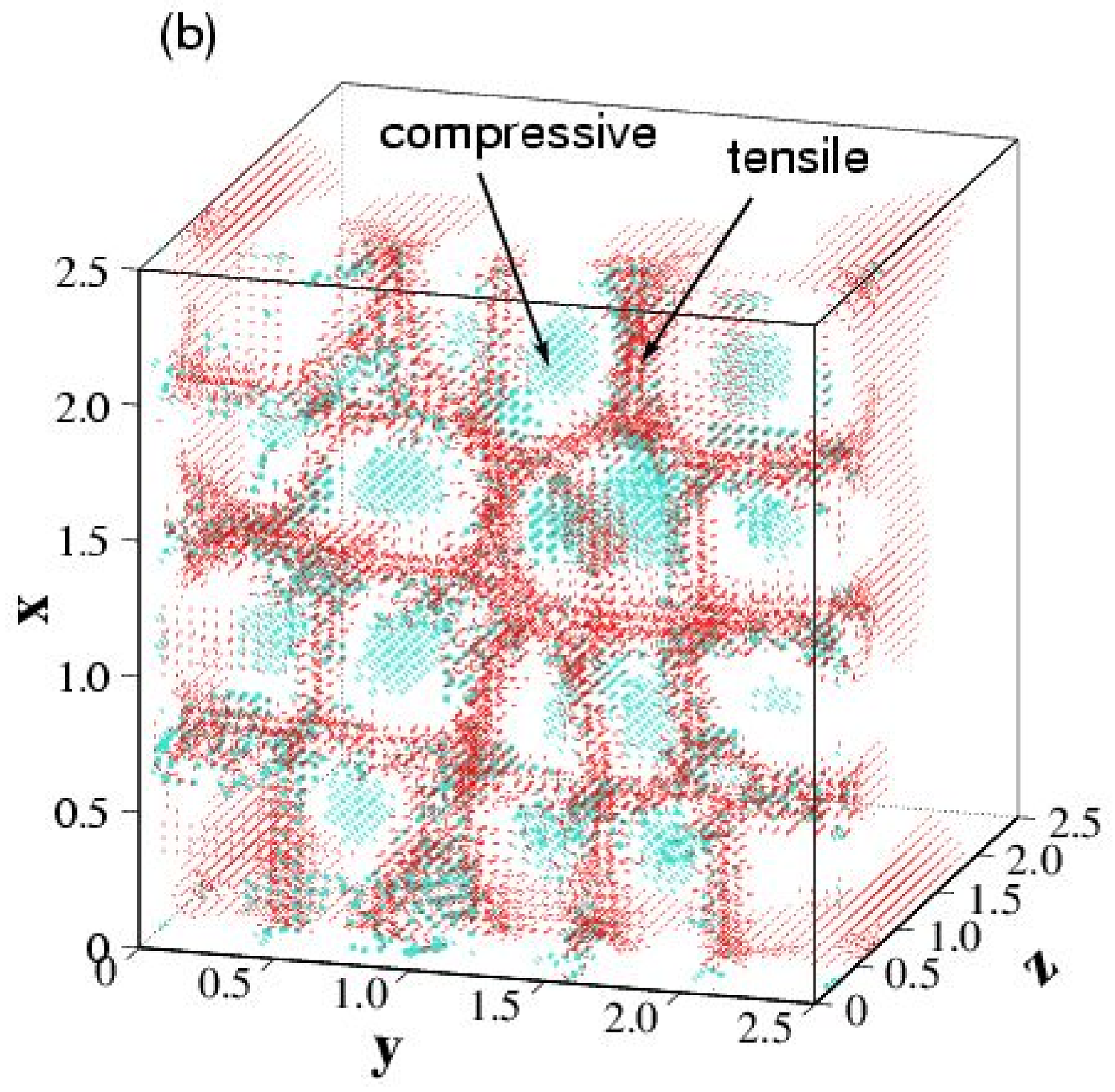}
\includegraphics[width=\psizel]{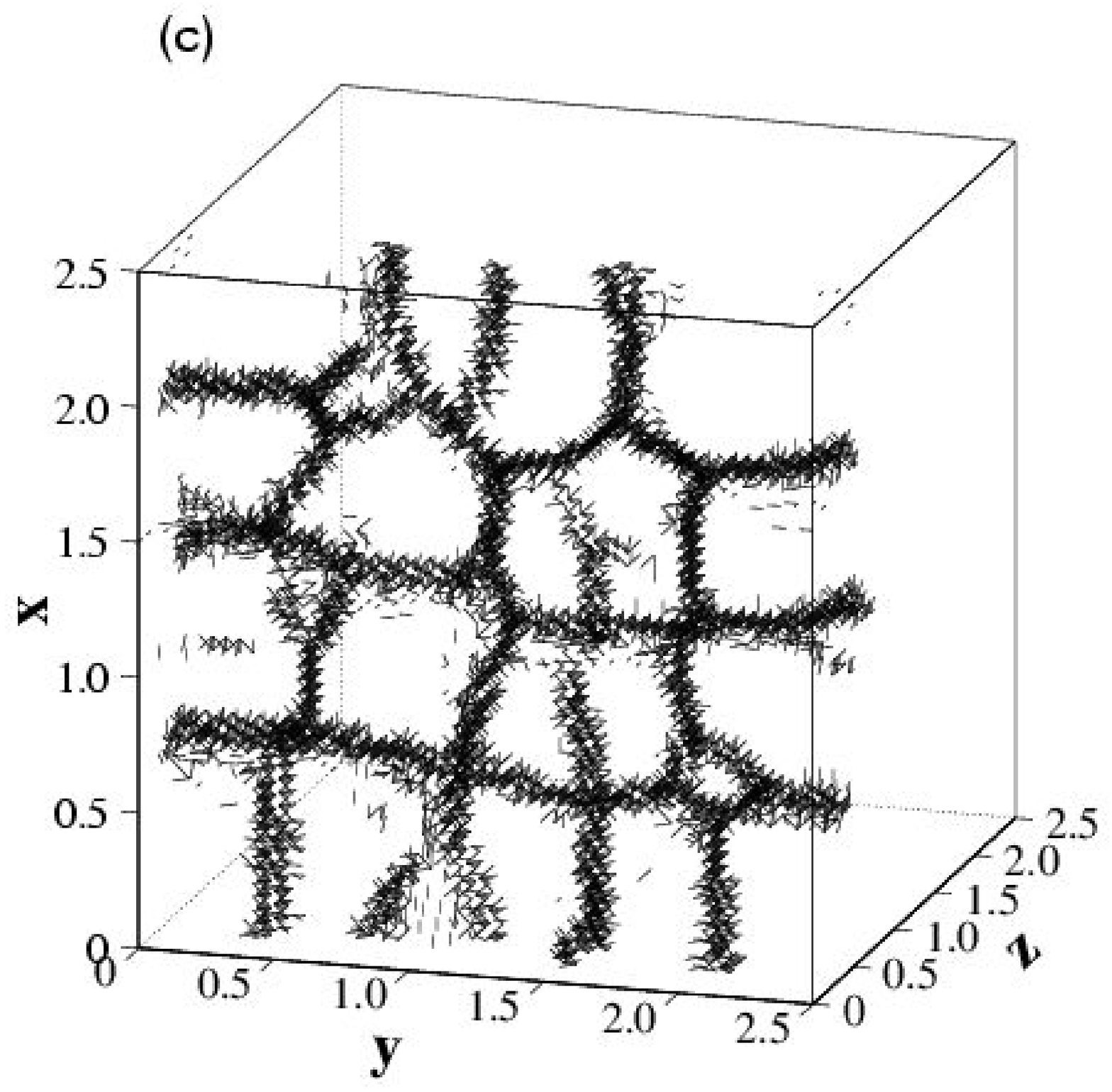}
\caption{
(Color online). Typical results of the numerical simulations.
The unit of axis is ${\rm cm}$.
(a)The water content field $\theta_L$ at $t=100\ {\rm h}$.
(b)The stress field at $t=100\ {\rm h}$.
The springs of which force is more than $0.4 F_c$ are plotted
in red\ (tensile) and blue\ (compressive),
or for grayscale image, in dark gray\ (tensile) and light gray\ (compressive).
In the front of cracks, the stress intensifies and the tensile springs accumulate.
Compressive springs accumulate between the crack tips.
(c)The crack pattern. The springs broken in $90<t<100\ {\rm h}$ are shown.
}
\label{typical}
\end{figure}

\begin{table}
\begin{center}
\begin{tabular}{ll}\hline
system size $L_x \times L_y \times L_z$ & $2.5\times2.5\times2.5\ {\rm cm^3}$\\
$N_x \times N_y \times N_z$ & $50\times50\times50$\\
density of liquid water $\rho_L$ & $1.0\times10^3\ {\rm kg\ m^{-3}}$\\
density of saturation vapor $\rho_{V0}$ & $1.7\times10^{-2}\ {\rm kg\ m^{-3}}$\\
saturation water content $\theta_S$ & 0.5\\
mass of a mol of water $M_w$ & $18\ {\rm g\ mol^{-1}}$\\
gas constant $R$ & $8.3\ {\rm J\ mol^{-1}\ K^{-1}}$\\
temperature $T$ & $293\ {\rm K}$\\
$RT/M_w$ & $1.35\times10^5\ {\rm J\ kg^{-1}}$\\
diffusion constant of vapor $D_V$ & $2.5\times10^{-5}\ {\rm m^2\ s^{-1}}$\\
tortuosity factor $\epsilon_V$ & $0.66$\\
evaporation rate $E_p$ & $0.01\ {\rm g\ cm^{-2}\ h^{-1}}$\\
atmospheric humidity $h_a$ & $0.5$\\
geometric mean particle diameter $d_g$ & $0.01 \sim 0.10\ {\rm mm}$\\
geometric standard deviation & $1.0$\\
of a particle diameter,\ $\sigma_g$ & \\ 
spring constants\ $k_1a_0$,\ $k_2a_0$ & $1.0$,\ $0.5$\\
critical force for the breaking $F_c$ & $0.003$\\
expansion coefficient $\kappa$ & $0.2$\\
time step $dt$ & $0.1\ {\rm h}$\\
\hline
\end{tabular}
\caption{Parameters of the water content field \cite{Campbell} and the stress field.}
\label{parameter_water}
\end{center}
\end{table}

The parameters used in the simulations are displayed in Table~\ref{parameter_water}.
The elastic and fracture parameters $k_1, k_2$ and $F_c$ are chosen as it works.
Below we show the results for the particle diameter $d_g=0.050\ {\rm mm}$ 
except for Figs.~\ref{water_s} and \ref{crack_d}.
The typical results at $t=100\ {\rm h}$ which corresponds to the middle of the stage 3
are shown in Fig.~\ref{typical}.
The water content decreases due to drying and
the region where the water content field changes sharply exists,
i.e., the drying front exists near $z \sim 0.4\ {\rm cm}$.
The non-uniform contraction near the drying front
consequently increases the stress in this region and fractures occur.
In the front of cracks, 
the stress intensifies and the tensile springs 
(denoted by red or dark gray in Fig.~\ref{typical}b) accumulate.
Compressive springs (blue or light gray) accumulate between the crack tips.
Figure~\ref{typical}(c) represents 
the springs broken between $t=90\ {\rm h}$ and $t=100\ {\rm h}$. 
The region where fractures occur actively forms a polygonal pattern
and the trace of such region develops into the prismatic structure of the cracks.
These results well capture the features observed in the real 
experiments \cite{Muller2,Toramaru,Goehring,guchi05,Goehring06}.
Next, we focus on the spatio-temporal behavior
of the key elements, i.e.,
the water content field, the stress field and the crack pattern.

\subsection{Water content field}
\begin{figure}\begin{center}
\includegraphics[width=\psizel]{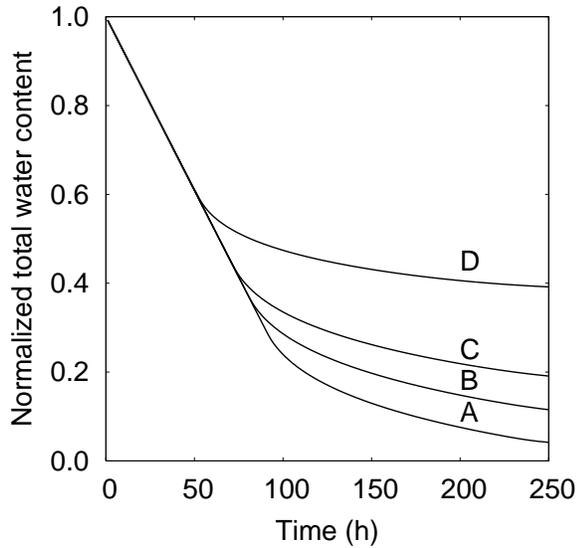}
\caption{Evolution of normalized total mass of liquid water$\int \theta_L(z,t)/\theta_S dz$.
A:\ $d_g=0.100\ {\rm mm}$,\ B:\ $d_g=0.050\ {\rm mm}$,\ 
C:\ $d_g=0.030\ {\rm mm}$\ and\ D:\ $d_g=0.010\ {\rm mm}$.}
\label{water_s}
\end{center}\end{figure}

\begin{figure}\begin{center}
\includegraphics[width=\psizel]{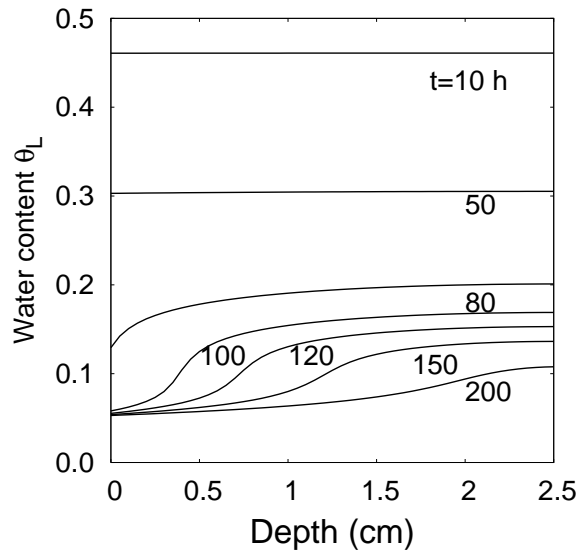}
\caption{Volumetric water content $\theta_L$ vs depth $z$.}
\label{water_t}
\end{center}\end{figure}

\begin{figure}\begin{center}
\includegraphics[width=\psizel]{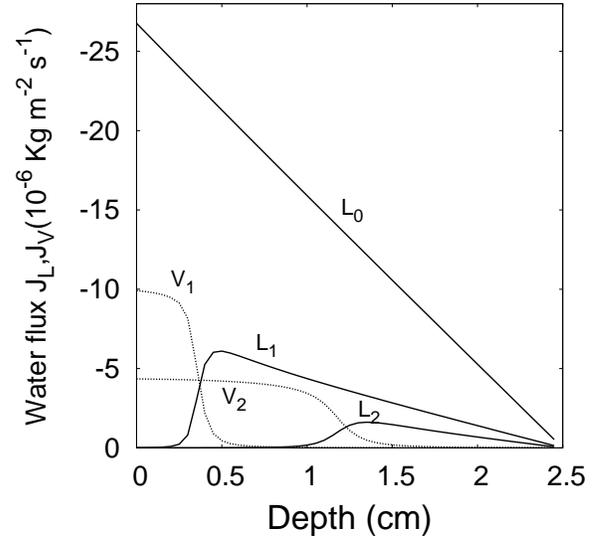}
\caption{Flux of liquid water $J_L$ and flux of vapor $J_V$ vs depth $z$. 
$L_0$:\ $J_L(t=50 {\rm h})$, $L_1$:\ $J_L(t=100 {\rm h})$, $V_1$:\ $J_V(t=100 {\rm h})$, 
$L_2$:\ $J_L(t=150 {\rm h})$ and $V_2$:\ $J_V(t=150 {\rm h})$.
$J_V(t=50 {\rm h})$ is on the order of $10^{-11}\ {\rm Kg\ m^{-2}\ s^{-1}}$.}
\label{water_j}
\end{center}\end{figure}

\begin{figure}\begin{center}
\includegraphics[width=\psizel]{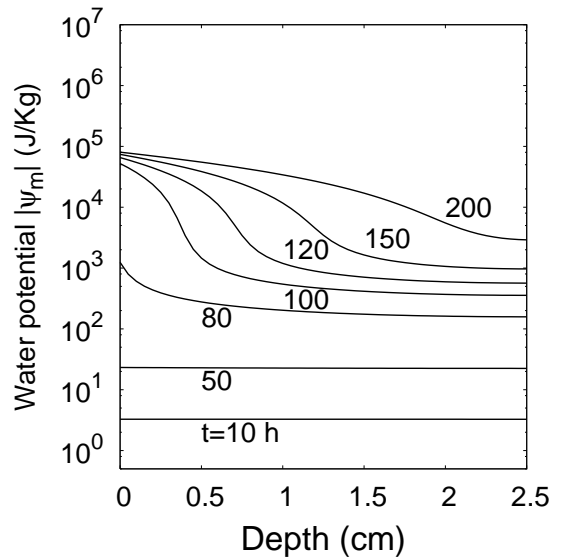}
\caption{Water potential~$|\psi_m|$ vs depth $z$. }
\label{water_p}
\end{center}\end{figure}

\begin{figure}\begin{center}
\includegraphics[width=\psizel]{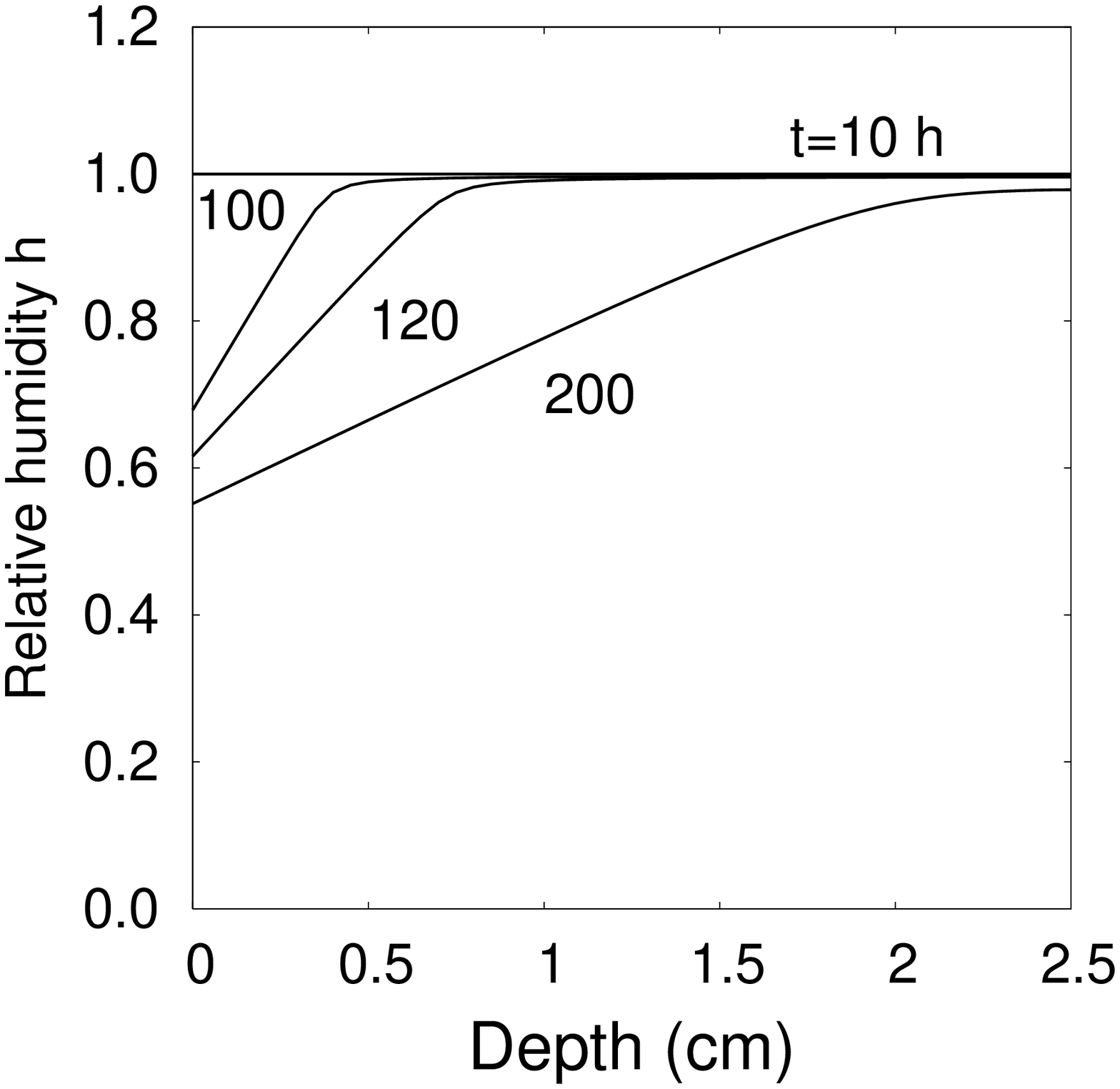}
\caption{Relative humidity $h$ vs depth $z$. }
\label{water_h}
\end{center}\end{figure}

The time evolution of the normalized total mass of liquid water 
$\int \theta_L(z,t)/\theta_S dz$ 
is displayed in Fig.~\ref{water_s}.
The two different stages are observed as in the experiments \cite{Muller2},
i.e., constant slope stage and slower drying stage.
The former and the latter correspond to the stage 1 and 3, respectively,
and the stage 2 connects them.
The time when the stage 1 switches to the stage 3
depends on the particle diameter $d_g$.
From the Kelvin eq.~(\ref{Kelvin}) and 
the relation between the water content and the water potential, Eq.~(\ref{eqnPm}) 
(shown in Fig.~\ref{water_param}),
as the particle diameter is large, the large amount of water evaporates.

The depth dependences of the several functions related to the water content field
are depicted in Figs.~\ref{water_t},\ref{water_j},\ref{water_p} and \ref{water_h}.
In the stage 1 ($0<t<50\ {\rm h}$), 
the volumetric water content $\theta_L$ decreases uniformly
and the flux of liquid water is dominant to the transportation of water.
In the stage 3 ($t>80\ {\rm h}$),
the drying front appears and propagates inward.
The dominant process of the transportation of water
differs between above and below the front, i.e.,
the flux of liquid water $J_L$ dominates below the drying front 
while the flux of vapor $J_V$ replaces it (Figs.~\ref{water_t} and \ref{water_j}).

Above the drying front,
the absolute value of the water potential $|\psi_m(z,t)|$
increases sharply to the order of $RT/M_w$($\sim 10^5\ {\rm J/kg}$)
and the relative humidity decreases below significantly $1$ (Figs.~\ref{water_p} and \ref{water_h}).
The contribution of the capillary water between micrometer-size particles
to the water potential
is on the order of $10^2\ {\rm J/kg}$.
So, the effects of the electrical and van der Waals force in the vicinity of the solid-liquid interface
are supposed to be important \cite{Iwata}.

\subsection{Stress field}
\begin{figure}
\includegraphics[width=\psizel]{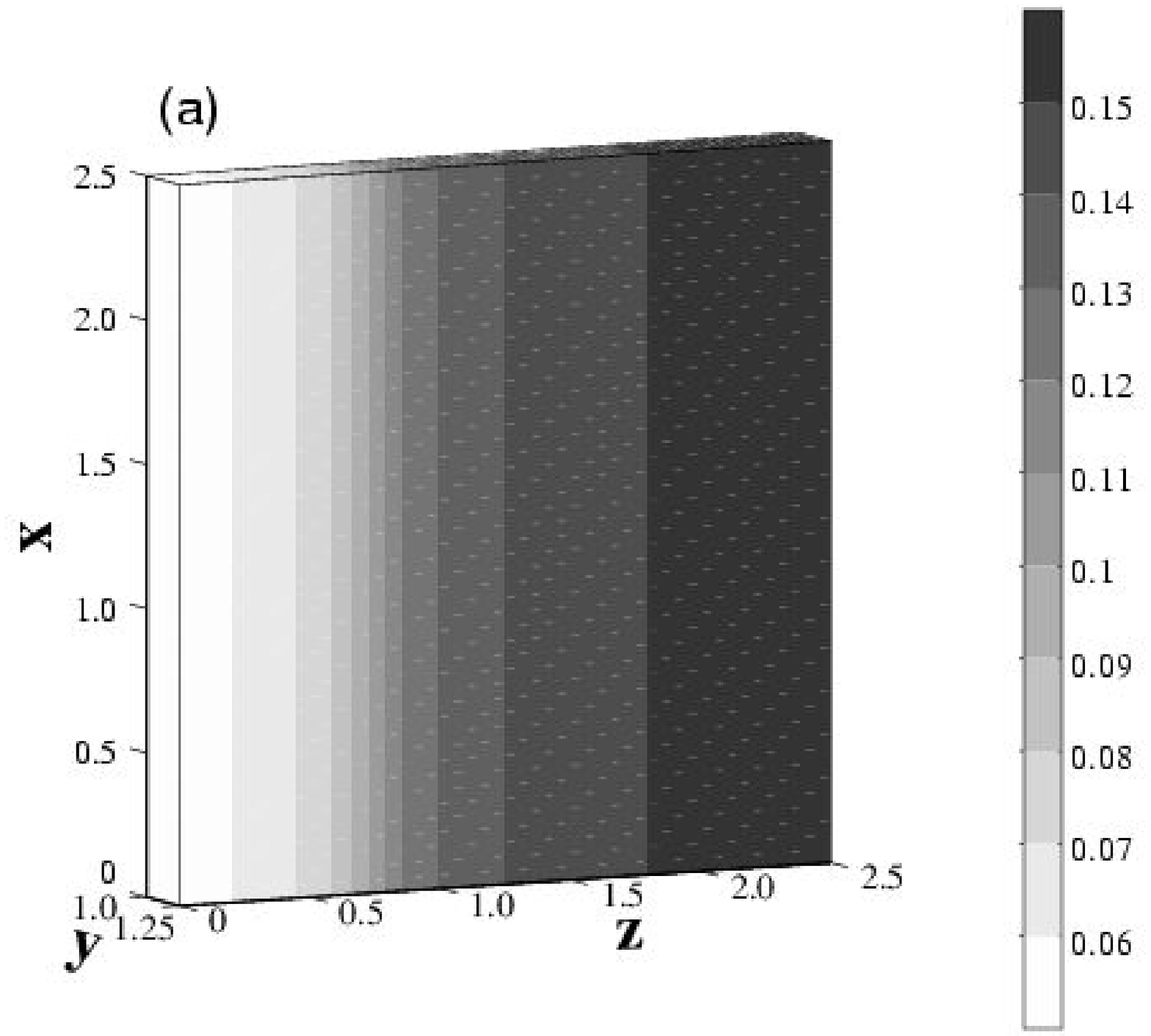}
\includegraphics[width=\psizel]{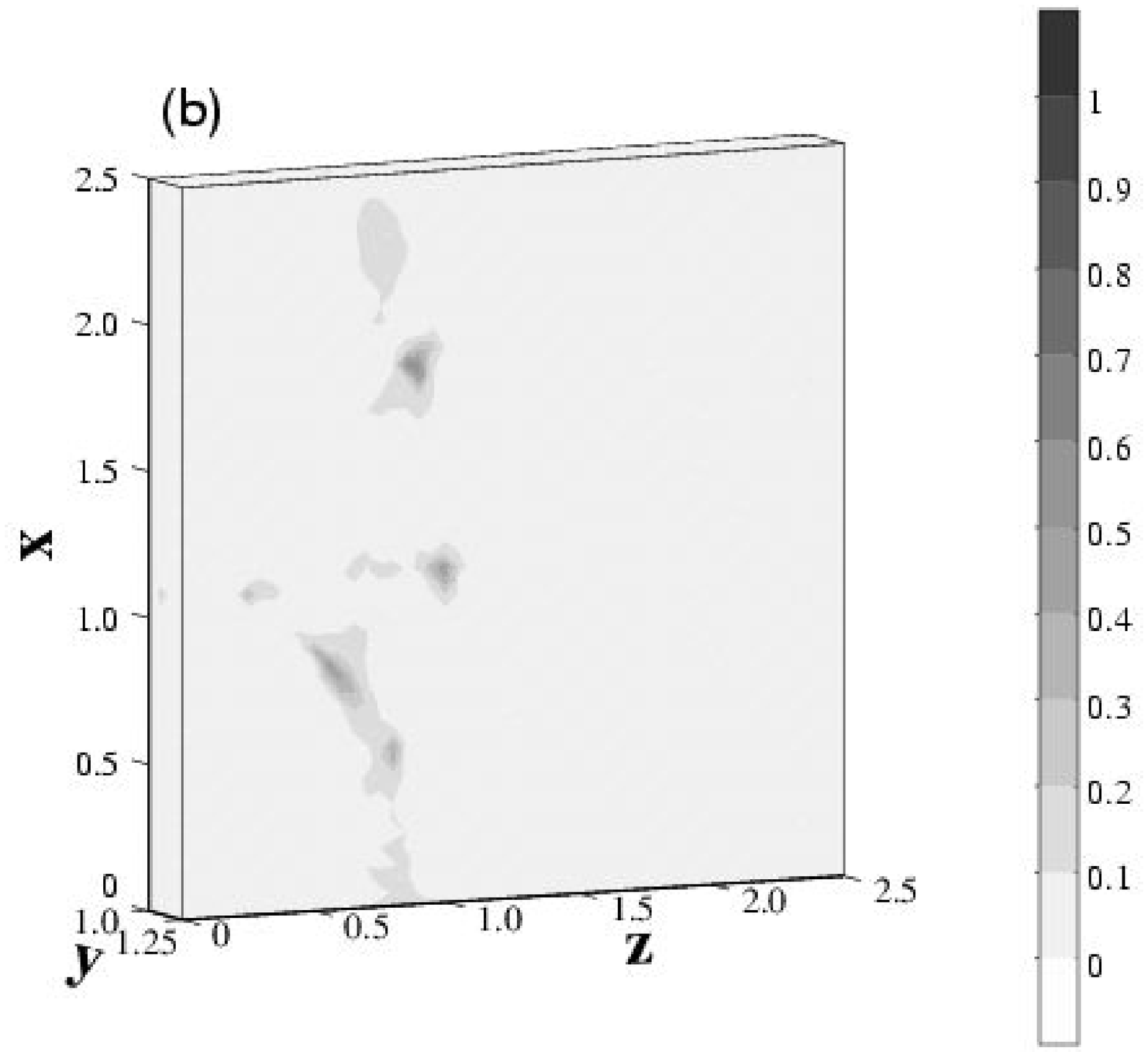}
\includegraphics[width=\psizel]{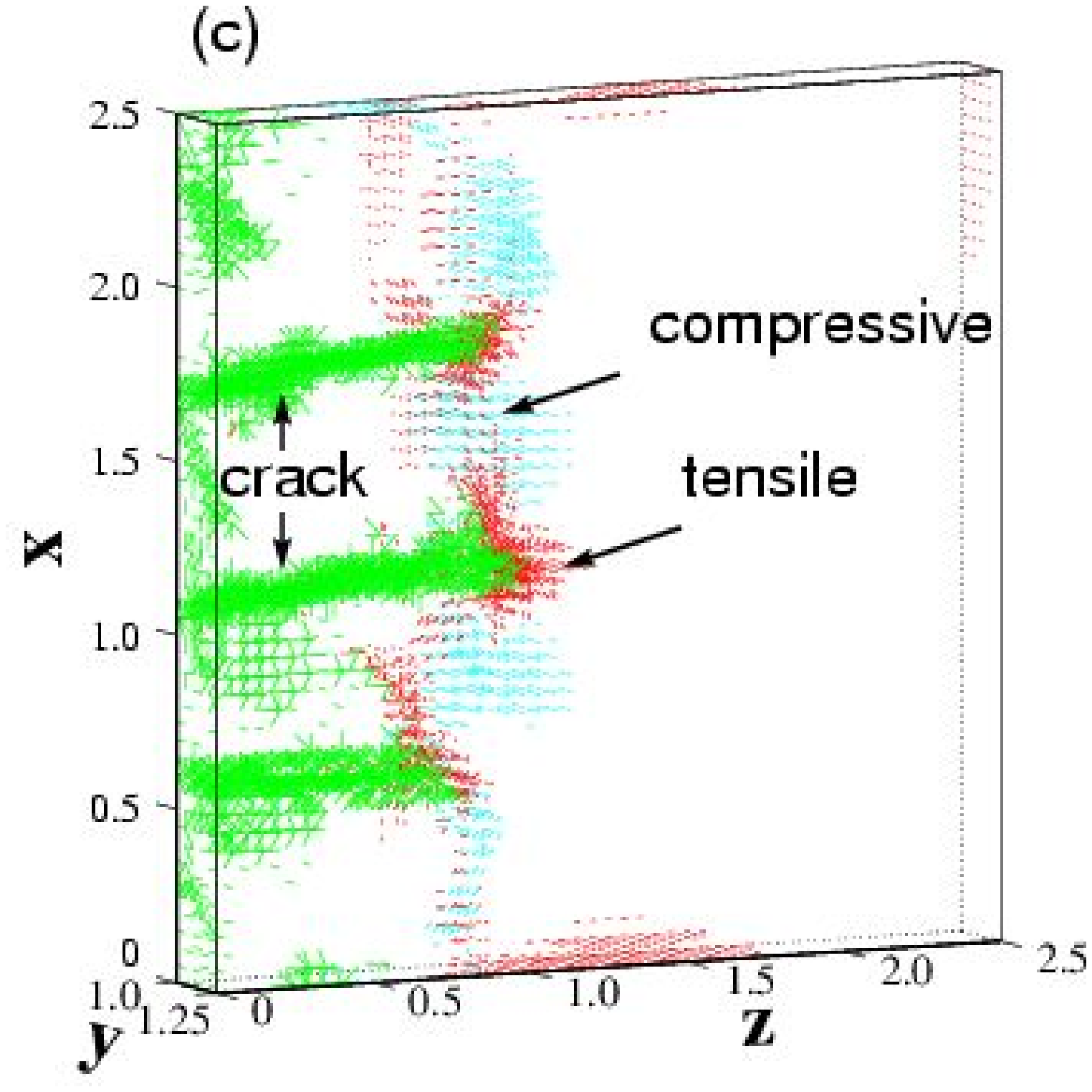}
\caption{
(Color online). The unit of axis is ${\rm cm}$.
(a)The water content field,
(b)the normalized elastic energy density,
and (c)the stress field are displayed in $1.0<y<1.25\ {\rm cm}$ at $t=120\ {\rm h}$.
The springs of which force is more than $0.3 F_c$ are plotted
in red\ (tensile) and blue\ (compressive),
or for grayscale image, 
in dark gray\ (tensile) and light gray\ (compressive).
The broken springs are plotted in green, or for grayscale image, in light gray.
In the drying front where the water content sharply changes,
the tip of cracks exists and the tensile springs accumulate.
Compressive springs accumulate between the crack tips.
}
\label{force}
\end{figure}

Figure~\ref{force} displays
the water content field, the normalized elastic energy density and the stress field 
on the vertical section of the system at $t=120\ {\rm h}$.
The energy density is normalized by the maximum value in the vertical section.
There are 
the drying front where the water content sharply changes
and the region where the tip of cracks exists
and the region where tensile springs (denoted by red or dark gray in Fig.~\ref{force}(c)) accumulate.
Compressive springs (blue or light gray) accumulate between the crack tips and
stress does not concentrate in the region far from the drying front.
The trace of the fractured region forms
the columnar crack pattern observed finally.
There is a crack of which tip does not follow to the drying front ($x \sim 2.0~{\rm cm}$).
It is an elementary process of columnar merger.

\subsection{Cracks}
\begin{figure}\begin{center}
\includegraphics[width=\psizel]{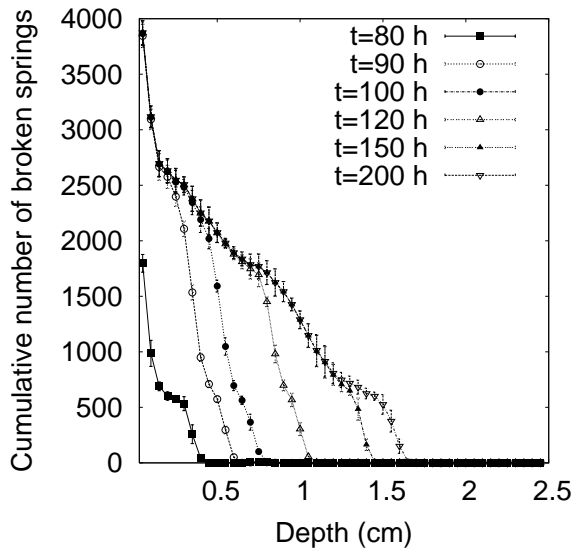}
\caption{Time evolution of the cumulative number of the broken springs at each horizontal cross sections.
The results obtained from seven different initial conditions
are averaged at a given depth $z$.}
\label{crack_t}
\end{center}\end{figure}

\begin{figure}
\includegraphics[width=\psizel]{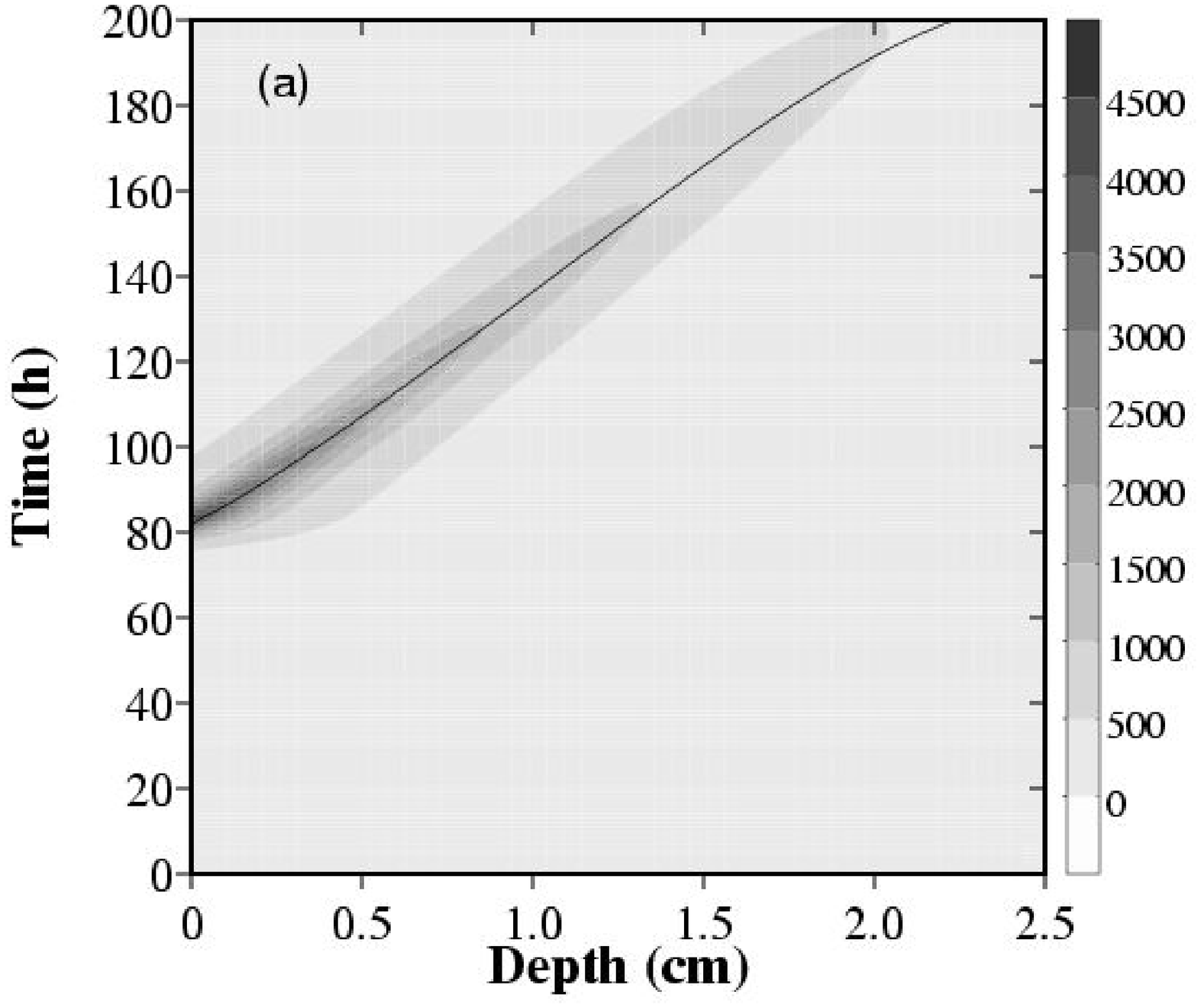}\\
\includegraphics[width=\psizel]{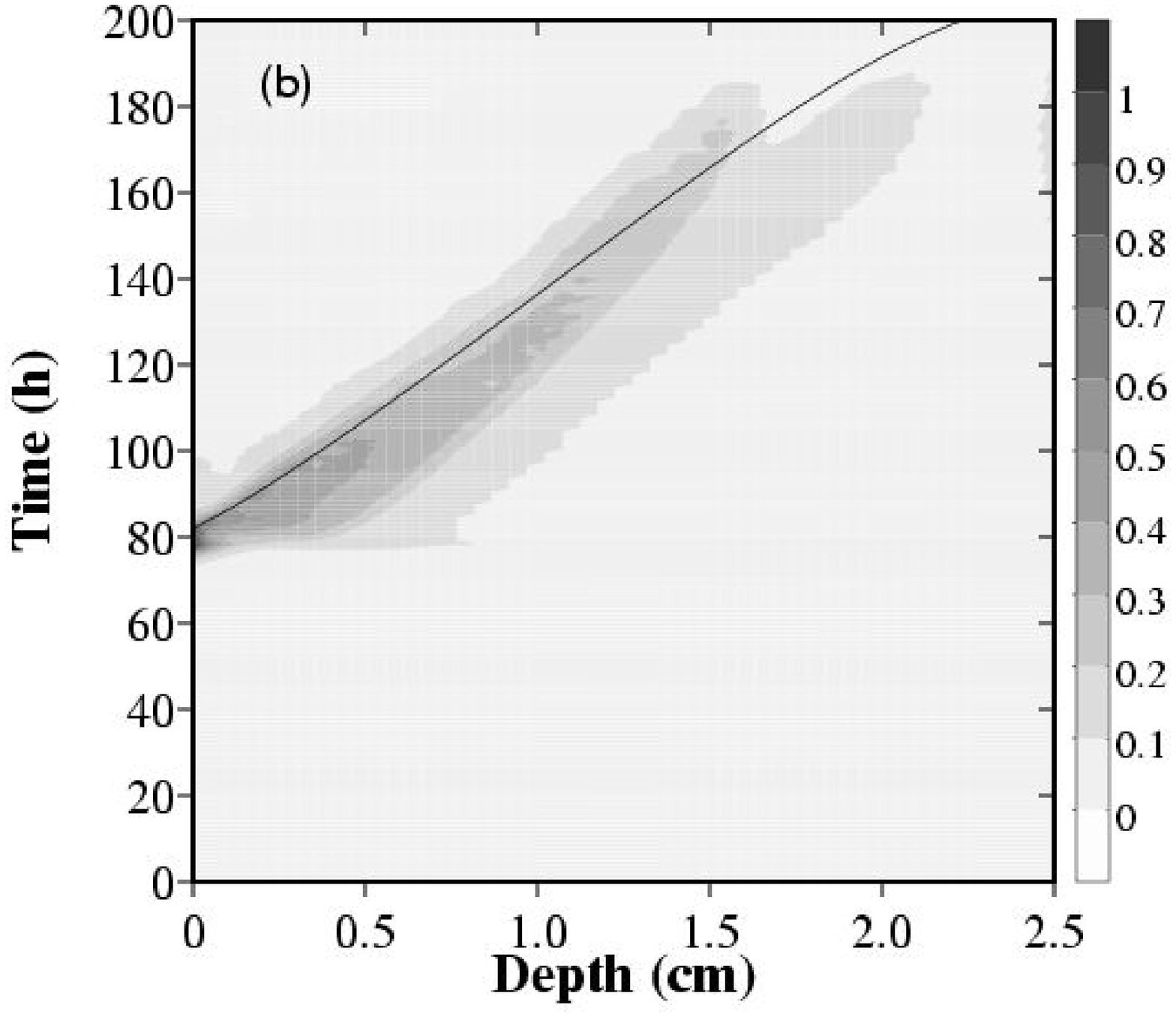}\\
\includegraphics[width=\psizel]{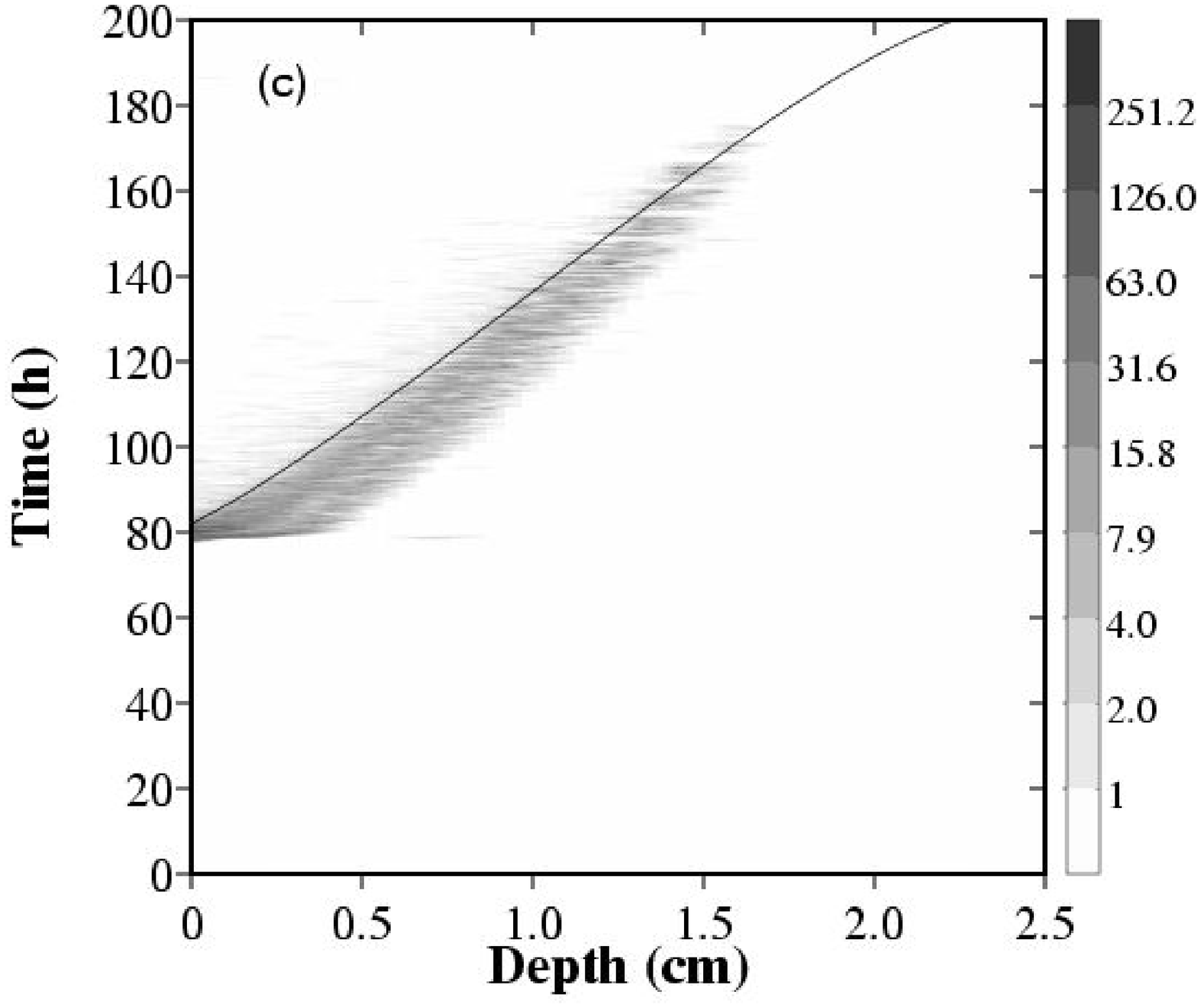}
\caption{Space-time plot of 
(a) $d\theta_L/dz$,
(b) the normalized elastic energy density,
and (c) the number of the broken springs.
These fields are averaged over the cross section at a given $z$.
The line indicates the depth
where the flux of vapor is equal to that of liquid water (see Fig.~\ref{water_j}),
which corresponds to the position of the drying front.}
\label{crack_water}
\end{figure}

\begin{figure}
\begin{tabular}{cccc}
\includegraphics[width=\psizes]{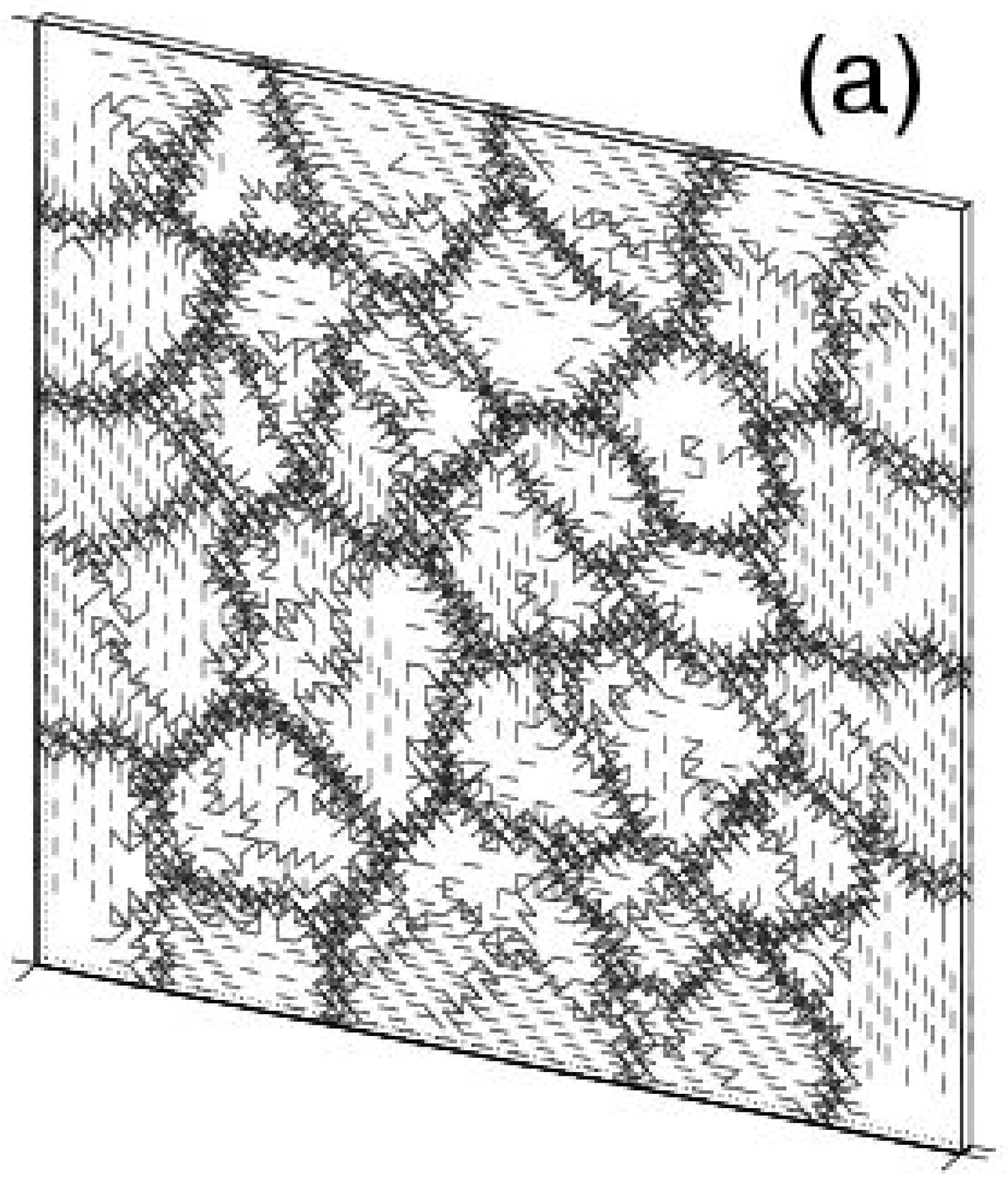}&
\includegraphics[width=\psizes]{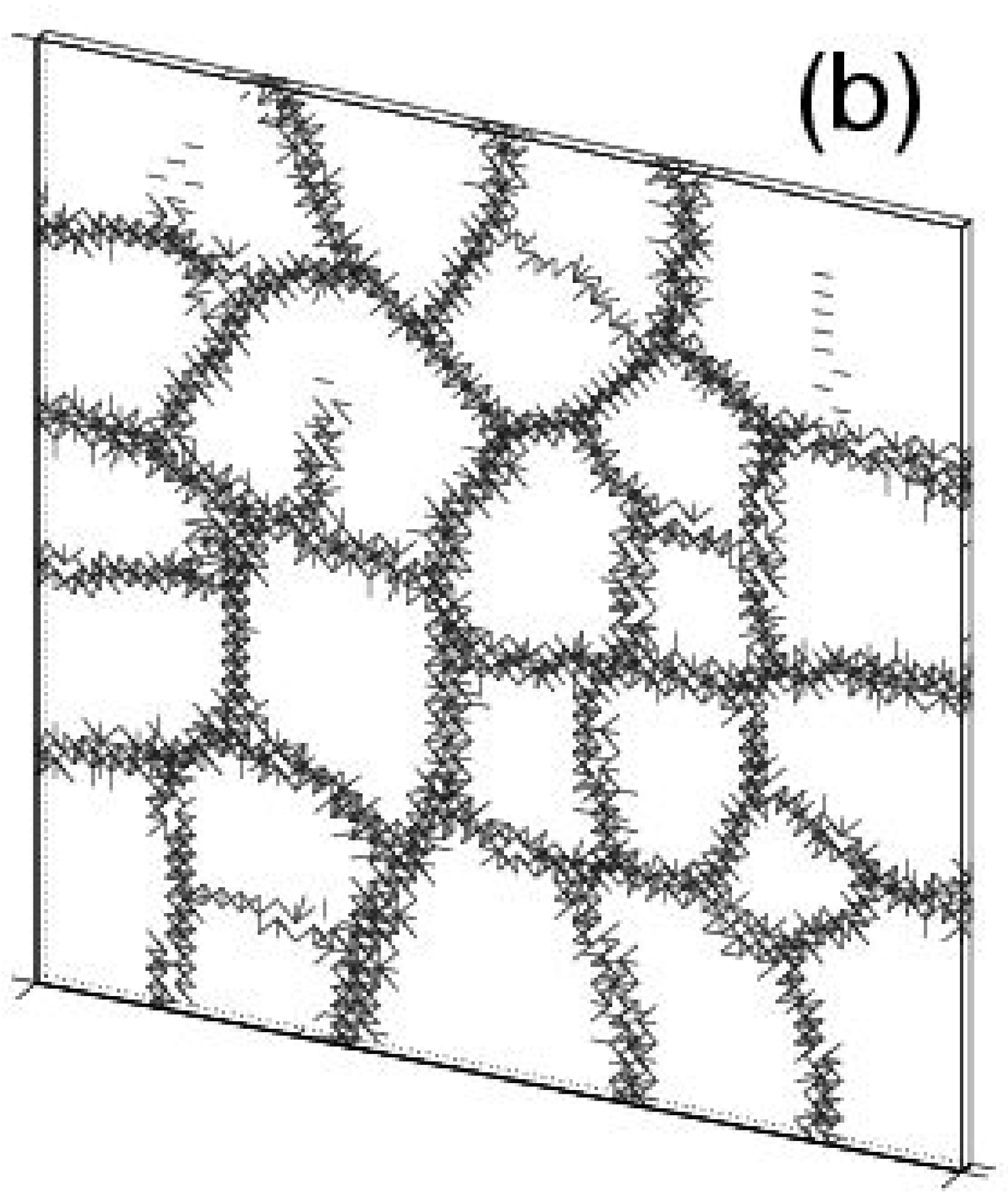}\\
\includegraphics[width=\psizes]{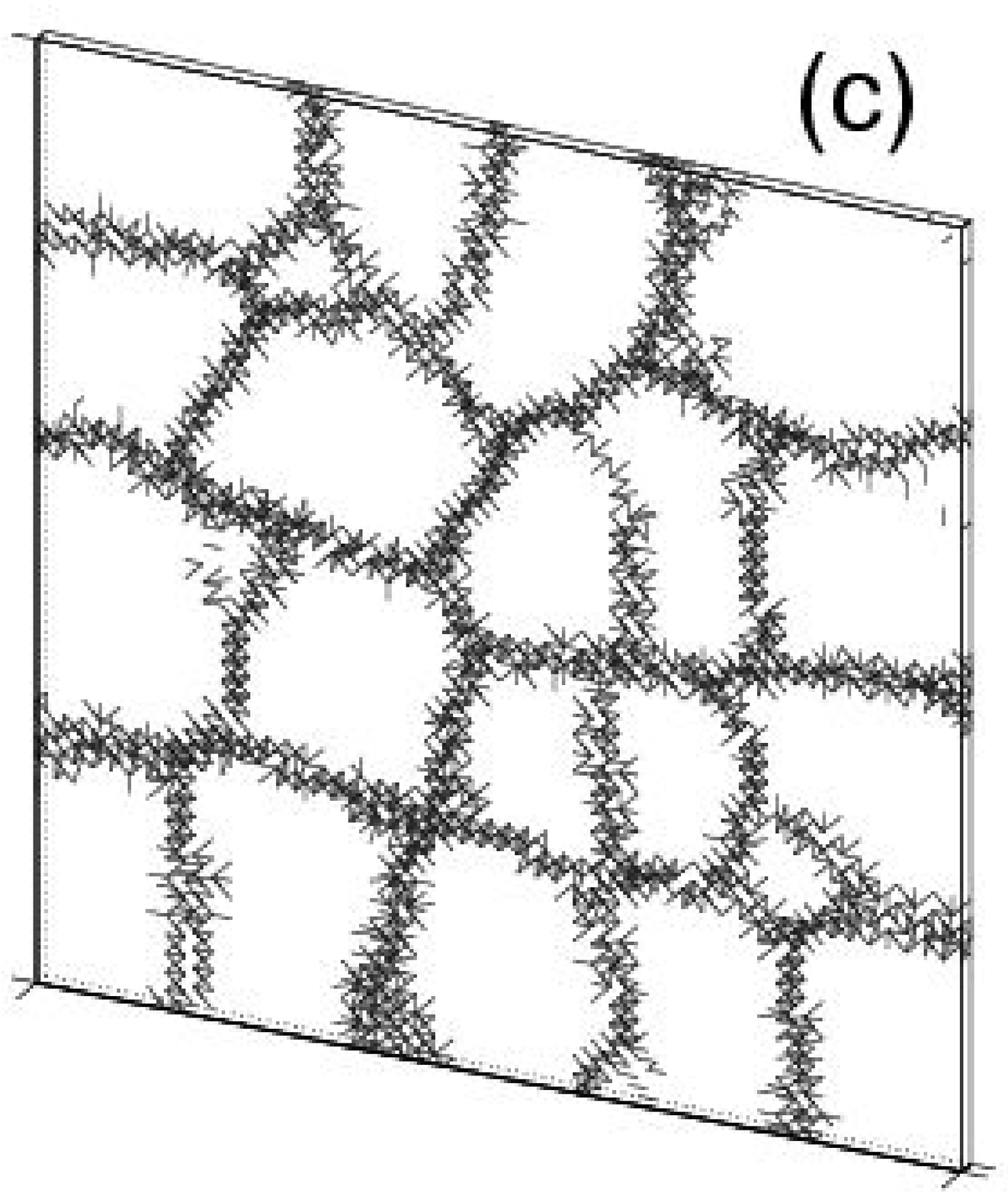}&
\includegraphics[width=\psizes]{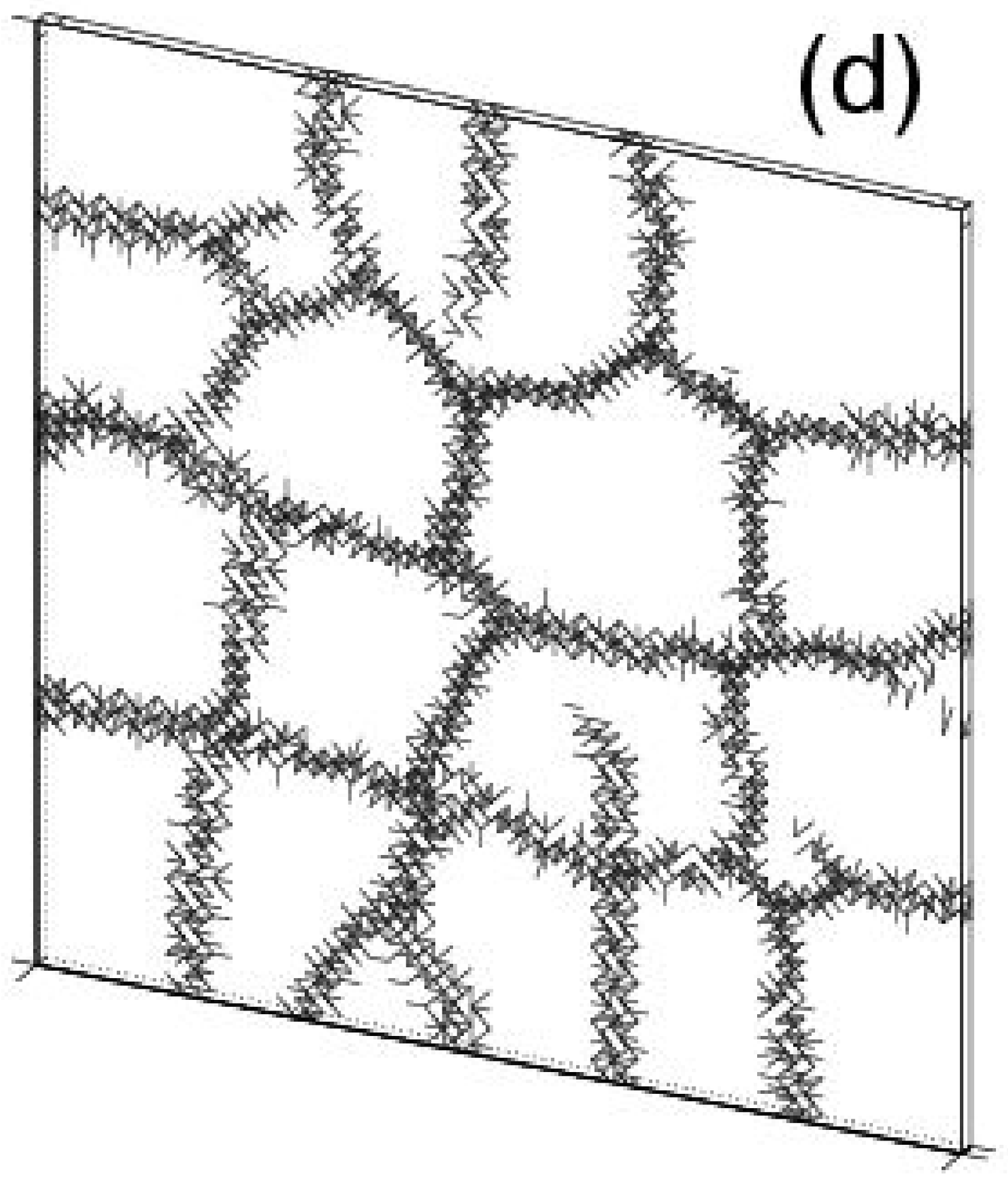}\\
\includegraphics[width=\psizes]{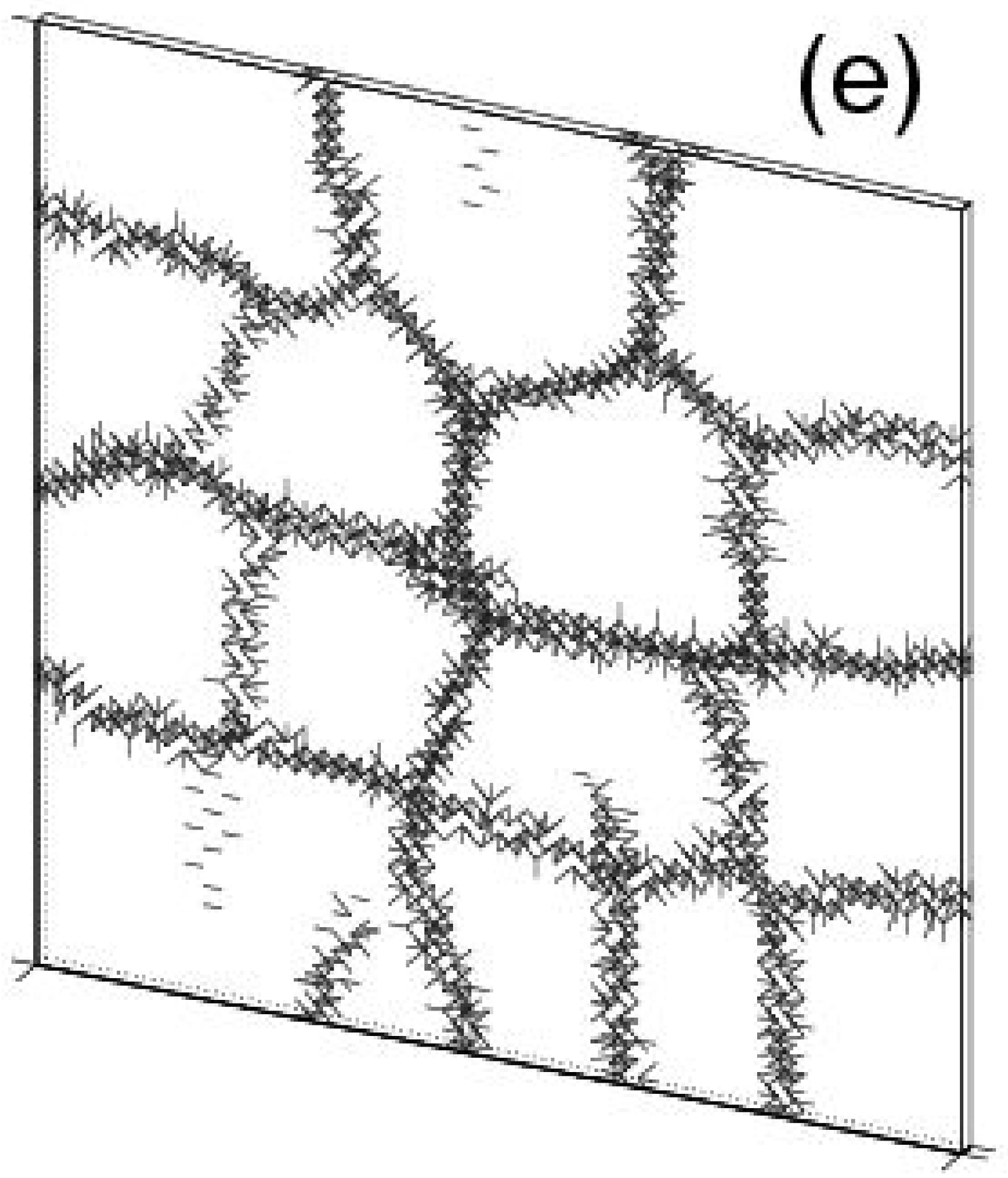}&
\includegraphics[width=\psizes]{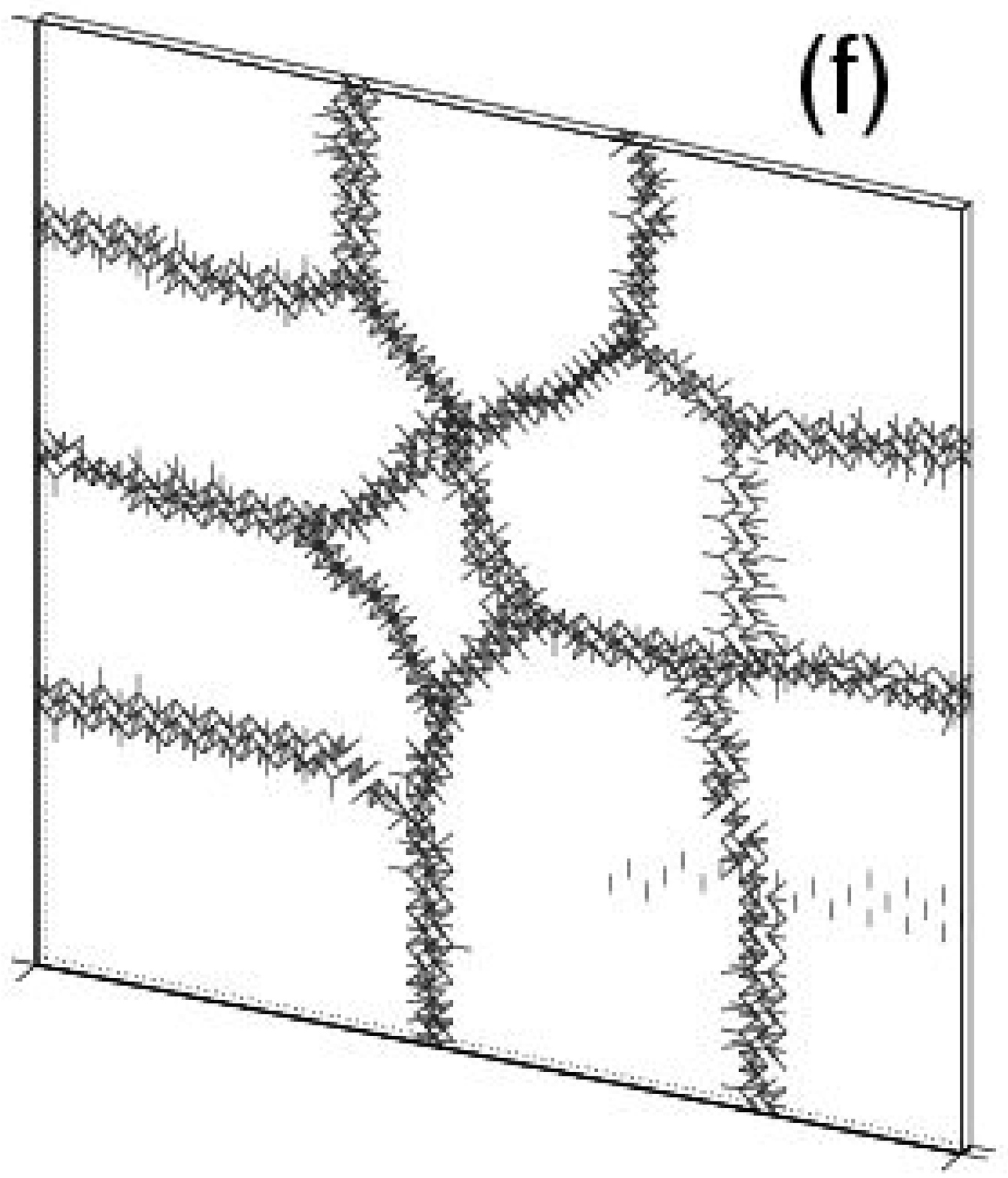}\\
\includegraphics[width=\psizes]{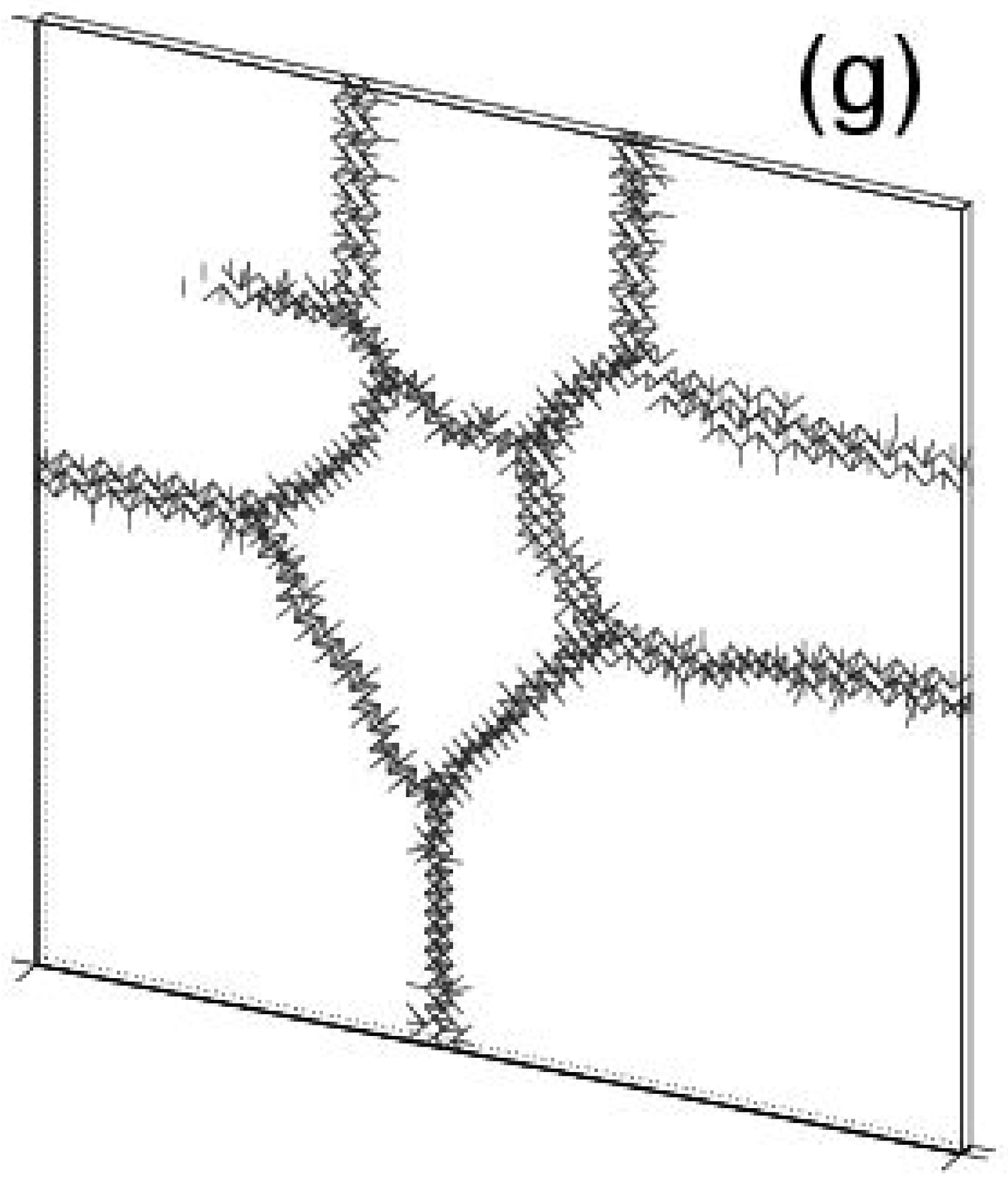}&
\includegraphics[width=\psizes]{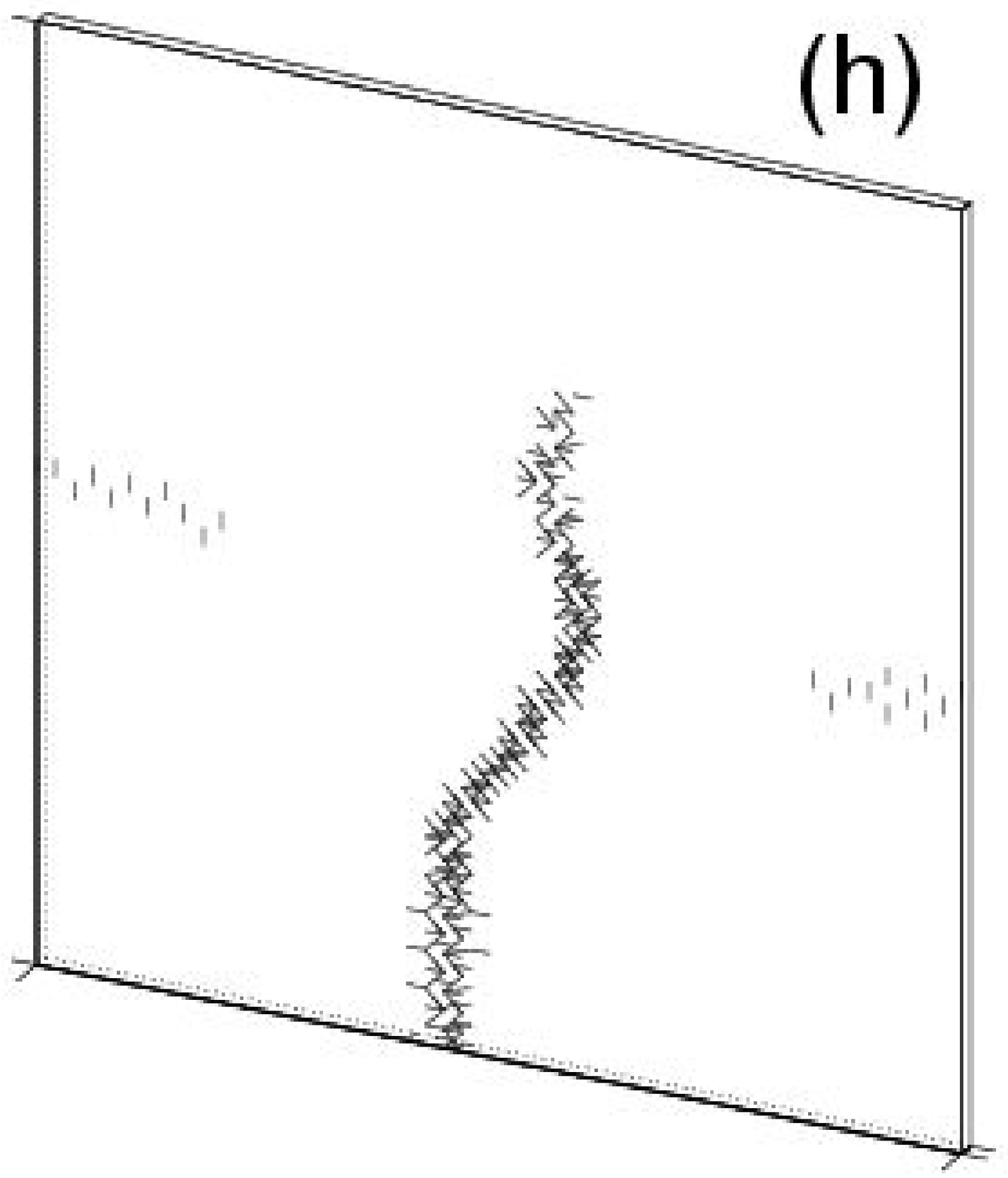}\\
\end{tabular}
\caption{Cross sections of the pattern of cracks at $t=200\ {\rm h}$.
(a)$z=0~{\rm cm}$,\ (b)$z=0.15~{\rm cm}$,\ (c)$z=0.3~{\rm cm}$,\ (d)$z=0.45~{\rm cm}$,\ 
(e)$z=0.6~{\rm cm}$,\ (f)$z=0.9~{\rm cm}$,\ (g)$z=1.2~{\rm cm}$\ and\ (h)$z=1.5~{\rm cm}$.
}
\label{crack_p}
\end{figure}

\begin{figure}\begin{center}
\includegraphics[width=\psizel]{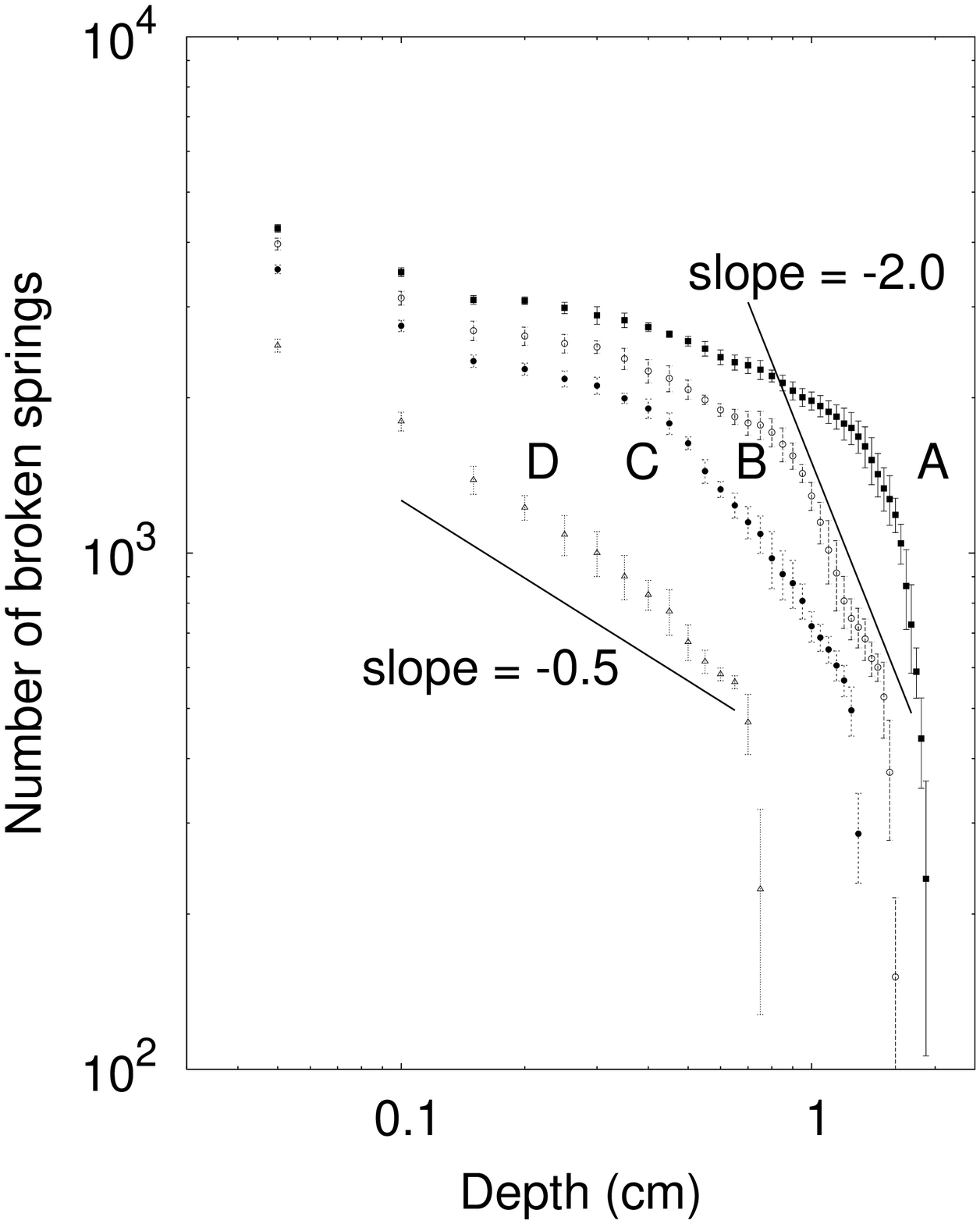}
\caption{
The cumulative number of the broken springs at each cross section 
at the final state ($t=200\ {\rm h}$)
for various particle diameters,
A:\ $d_g=0.100\ {\rm mm}$,\ B:\ $d_g=0.050\ {\rm mm}$,\ 
C:\ $d_g=0.030\ {\rm mm}$\ and\ D:\ $d_g=0.010\ {\rm mm}$.
The results obtained from seven different initial conditions
are averaged for each particle diameter.
}
\label{crack_d}
\end{center}\end{figure}

Cracks are formed when the uniformity of the water content distribution is lost
and the tips of cracks propagate inward.
The time evolution of $\sum_{\tau<t} l(z,\tau)$ are shown in Fig.~\ref{crack_t},
where $l(z,\tau)$ represents the number of springs broken at $(z,\tau)$.
$\sum_{\tau<t} l(z,\tau)$ corresponds to the cumulative crack length at $(z,t)$.
The area surrounded by the two curves at different times
corresponds to the number of springs broken between those times.
From the fact that the difference between the distribution of broken springs
for two adjacent time has finite support,
the active cracking area is localized in $z$-direction.
The relation between the position of active cracking area 
and the drying front can be checked as follows.
The space-time plots of 
$d\theta_L(z,t)/dz$, $\sum_{X,Y} E_{\alpha}$ and $l(z,t)$, 
are shown as gray scale images in Fig.~\ref{crack_water}.
The energy density is normalized by the maximum value in all region.
It is confirmed that the active cracking area moves inward
following the drying front.
However, the number of the broken springs is not the simple function of 
the water content $\theta_L(z,t)$ or its derivative.

The typical patterns of cracks formed up to the final state ($t=200\ {\rm h}$)
are shown in Fig.~\ref{crack_p}.
The polygonal columnar structure can be seen 
despite of the anisotropy of the cubic lattice.
In ref.~\cite{Goehring06}, two mechanism for changing scale are suggested, 
i.e., merging and column initiation from a vertex. 
In Fig.~\ref{crack_p}, both process are observed. 
These features well capture the experiments of 
starch-water mixtures \cite{Muller2,Toramaru,Goehring,guchi05,Goehring06}.
Now, we analyze the depth dependence of the final crack pattern.
Figure~\ref{crack_d} exhibits the depth dependence of the cumulative crack length
up to the final state with double logarithmic plot
to compare with the experimental data.
In contrast to the real experiments,
we can easily change the physical parameters.
If we change the particle diameter $d_g$,
it affects the $\psi_{m0}$ through phenomenological Eq.~(\ref{eqndg}).
The size of columns decreases as the particle diameter increases.
Though it is difficult to state definitely,
there is a regime in which the number of broken springs decreases with a power law
with exponent between $-0.5$ and $-2$.

%
%
\section{Discussions}
In this section, we note several points such as
the translation of the evolution equation
from the matric potential $\psi_m(z,t)$ to the water content $\theta_L(z,t)$,
the drainage effect of cracks,
the geological columnar joints,
the feedback from the stress field to the water content field,
the characterization of the crack patterns and
the properties of the granules water mixtures.

\begin{figure}\begin{center}
\includegraphics[width=\psizel]{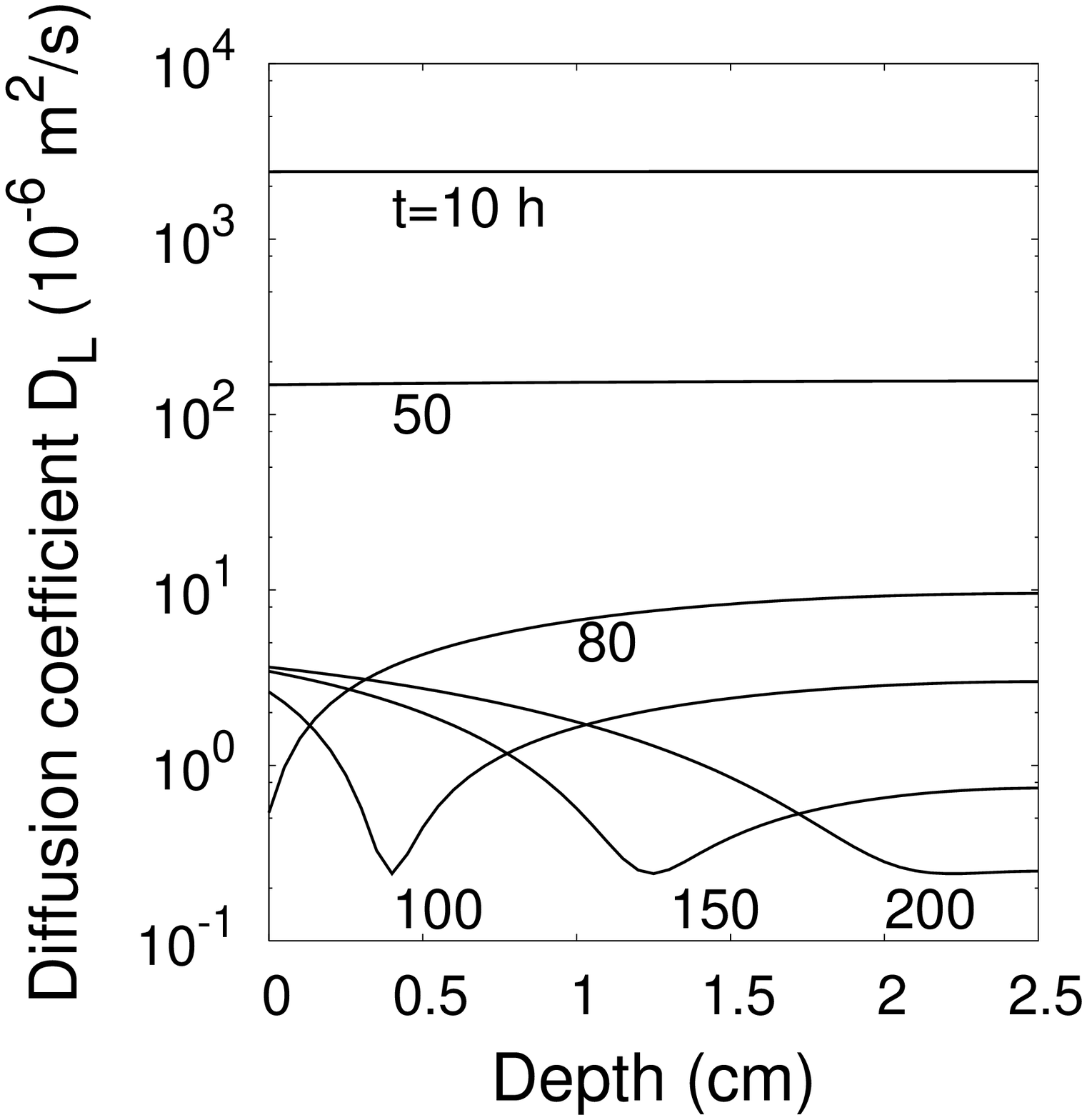}
\caption{Diffusion coefficient $D_L$ defined by Eq.~(\ref{eqnDL}).}
\label{water_d}
\end{center}\end{figure}

First, we discuss the drying front which is obtained
from the evolution equation of the water potential (Eq.~(\ref{eqnP})).
Changing the variable from $\psi_m(z,t)$ to $\theta_L(z,t)$ by Eq.~(\ref{eqnPm}),
we get
\begin{equation}
\rho_L\{1+g(\theta_L)\} \frac{\partial \theta_L(z,t)}{\partial t}=
\frac{d}{dz}
\left( D_L(z,t) \frac{d}{dz} \theta_L(z,t)\right), \label{eqnC}
\end{equation}
where
\begin{eqnarray}
g(\theta_L)= 
-\frac{\rho_V(z,t)}{\rho_L}\left(
1+\frac{b\psi_m(z,t) M_w \theta_G(z,t)}{RT \theta_L(z,t)}\right),\\
D_L(z,t)=\frac{-b\psi_m(z,t) (k_L(z,t)+k_V(z,t)) }{\theta_L(z,t)}. \label{eqnDL}
\end{eqnarray}
$g(\theta_L)$ is of order $10^{-4}$ due to $\rho_V \ll \rho_L$;
therefore, if we ignore $g(\theta_L)$,
the time evolution of $\theta_L$ can be considered to obey 
the nonlinear diffusion equation with variable diffusion coefficient $D_L(z,t)$.
In our simulations the diffusion coefficient $D_L(z,t)$ has a minimum near the drying front (Fig.~\ref{water_d})
as Goehring et al. \cite{Goehring06} pointed in their experiments.
The nonlinearity of the diffusion coefficient is crucial for the water content distribution
which has a concave regime.

Second, we discuss the assumption that cracks do not affect the water content field.
Assuming the crack aperture width $l_c \sim 10^{-1}\ {\rm mm}$
and the propagation speed of the drying front $v_w \sim 10^{-5}\ {\rm mm/sec}$
and the volumetric gas content near the drying front $\theta_G \sim 0.4$,
the length of the vapor diffusion in $\tau_w=l_c/v_w$ is 
on the order of $\sqrt{D_v \epsilon_V \theta_G \tau_w}$ $\sim 10^2\ {\rm mm}$.
Because this is much larger than 
the crack aperture width,
the density of vapor in a crack aperture
is determined quasi-statically
by the water potential on the crack surface.
Therefore, the effect of cracks on the dynamics of the water content field as the boundary condition
can be negligible
above the drying front because the flux of vapor is dominant in this region.
The flux of liquid water may be affected by the cracks
in the region where the two fluxes are comparable.
However we infer that the flux of water does not change significantly
because the region is limited in the vicinities of the crack tips.
Hence the assumption is valid except for the small effect near the drying front.

For studying the effect of cracks on the water content field,
we also perform numerical simulations based on the different assumptions
that the transportation of liquid water obeys the linear diffusion equation
and the diffusion constant in cracks is larger than in medium.
The result, however, is that
the broken springs do not form planes to become crack surfaces
but form some clusters
and no stable prismatic pattern is observed.
Recently, S{\o}renssen et al. \cite{Sorenssen} report a similar fracture simulation
under the condition that
the quantity which determines the local volume reduction
diffuses to the boundary of the solid
and find that the zone where the fracture occurs actively propagates
and the fracture patterns resemble a forest of trees, 
i.e., there is many branches of cracks.
In order to obtain the prismatic fracture pattern 
under the condition that the water content field does not affect the stress field
except for determining the volume shrinkage rate,
the mechanism that
the external field forms the front by itself without cracks
would be necessary.
It is the nonlinear diffusion of water in our simulations.

Third, let us consider the another columnar joint formation phenomenon, i.e.,
the geological columnar joint formation driven by lava cooling process.
It is accepted that the cooling process of lava is the combination of 
the linear diffusion of heat and the convective heat loss through cracks
\cite{Hardee,Reiter,Degraff89,Degraff93,Budkewitsch}. 
On the other hand, in our spring network model
we obtained the prismatic structure of cracks
in the condition that cracks do not affect the water content field and 
the water pressure do not affect the stress field. 
There is the difference in the external field, i.e.,
the thermal field above the front is nearly constant \cite{Hardee} 
and the water content field below the front is nearly constant \cite{guchi05,Goehring06}.
So it is a future work to study numerically the cooling joints 
with taking account of the convective heat loss through cracks 
and the external stresses i.e., the horizontal tectonic stress, 
the vertical overburden stress and the local pore pressure.

Fourth, we discuss the feedback from the stress field to the water content field.
We ignore this feedback, i.e., 
the water potential is assumed not to be affected by the change of the stress field.
In our simulations, as described in Sec.~III~B,
the contribution of the capillary water pressure to the water potential
becomes much smaller than the absolute value
when the water content decreases.
Therefore, we infer that,
when the drying front appears,
the deformation does not change significantly the water potential
through the water pressure.
This inference is consistent with our assumption.
However it is possible that the deformation may affect
the properties of water on the particle surfaces and the shapes of the water-vapor interfaces.
It is a future subject to investigate the contribution of these effects to the water potential.
%

Fifth, we discuss the characterization of the crack patterns.
The fracture occurs near the propagating front of the water content field.
The trace of the fractured region forms the columnar pattern with polygonal cross sections.
The size of columns increases with depth.
In this paper, as a key quantity, we focus on the crack length
(i.e., the number of broken springs) which is supposed to be proportional
to the surface energy in connection with the Griffith theory \cite{Griffith}.
Other characterizations, such as the mean polygon area and number of polygon sides,
seems to be worth analyzing as a future work. However, they should require larger
system size and careful definition of the quantities. 
If the pattern at the cross section is assumed to be isotropic,
the characteristic length of the pattern is inversely proportional to the crack length,
so it is possible to compare the depth dependence of the cell area at the cross section
with the experimental results.
We suggest that the drying granules-water mixtures of larger particles forms the smaller size of columns.
the dependence of the size of columns on the particle diameter 
is a future subject.

Finally, we discuss the properties of the granules water mixtures.
We obtain the prismatic pattern of cracks
by employing the relations for general soils in our simulations.
For the present, it is known that
only the combination of starch and water forms the prismatic fracture pattern
among drying mixtures of various combinations of granules and liquids.
For understanding why dried mixtures of other granules and water do not form the columnar structure,
it is necessary 
to measure experimentally various material properties such as
the expansion coefficient
and the dependences of the water potential and the hydraulic conductivity 
on the water concentration.

%
%
\section{Conclusions}
We study the drying process and the columnar fracture process
in granules-water mixtures.
The mixture is assumed to be the elastic porous medium
and each process is represented by 
a phenomenological model 
with the water potential and a spring network, respectively.
We assume that
the cracks do not affect the water content field as a boundary condition
and ignore the feedback from the stress field to the water content field.
In the numerical simulations,
we find that
the water content distribution has a propagating front
and the pattern of cracks exhibits a prismatic structure.

\begin{acknowledgments}
We acknowledge 
Y. Sasaki, 
I. Aoki,
Y. Kuramoto, 
S. Shinomoto,
H. Nakao 
and all members of the nonlinear dynamics group, Kyoto University
for fruitful discussions and encouragements.
The numerical calculations were carried out on Altix3700 BX2
at YITP in Kyoto University.
\end{acknowledgments}

\bibliography{Nishimoto}

\begin{thebibliography}{44}
\bibitem[{typ()}]{type12}
\bibinfo{note}{{The names `primary' and `secondary' cracks seem to imply the
  temporal order of the phenomena. In some cases, however, only the "secondary"
  crack forms under specific conditions. Therefore, more neutral names such as
  `type I' and `type II' \cite{guchi05} may be adequate.}}
\bibitem[{suc()}]{suction}
\bibinfo{note}{{The water potential has several aliases such as suction, head
  and soil-water potential.}}
\bibitem[{eva()}]{evaporation}
\bibinfo{note}{{In order to simulate the constant evaporation rate devised in
  \cite{Goehring06}, feedback control is required.}}
\end{thebibliography}

\end{document}